\documentclass[aps, prl, superscriptaddress, reprint, twocolumn]{revtex4-2}
\usepackage[T1]{fontenc}
\usepackage[utf8]{inputenc}

\usepackage{amsmath, amssymb, amstext, amsfonts, amsthm}
\usepackage{mathtools}
\usepackage{dsfont}
\usepackage{graphicx}
\usepackage[dvipsnames]{xcolor}
\usepackage{float}
\usepackage{ragged2e}
\usepackage[caption=false, justification=justified]{subfig}
\DeclareCaptionJustification{justified}{\justifying}
\usepackage{placeins}
\usepackage{standalone}
\usepackage{tikz}
\usepackage{enumitem}
\usepackage{microtype}

\usepackage{stix2}    %And stix2 is about equally good as libertine, but has a more formal flavor to it that may be more appropriate for scientific documents. Maybe will go with stix2.
\usepackage{subdepth} %makes subscripts appear at the same heights independent of whether the variable has a superscript
\usepackage[colorlinks=true, linkcolor = BlueViolet, citecolor = BlueViolet, urlcolor = BlueViolet]{hyperref}

\theoremstyle{definition}

\DeclareMathOperator{\Tr}{Tr}

\newcommand{\idop}{\mathds{1}}  %note that the hat has been removed for now
\newcommand{\dos}{\mathcal{N}}

\newcommand{\htime}{{\rm H}}

\newcommand{\epr}{\mathrm{EPR}}

\newcommand{\ramp}{\text{ramp}}

\newcommand{\edit}[1]{{#1}} %for revisions

\begin{document}

\title{Enhanced entanglement from quantum ergodicity}
\author{Amit Vikram}
\affiliation{JILA and Center for Theory of Quantum Matter, Department of Physics, University of Colorado, Boulder CO 80309 USA}

\begin{abstract}
    The quantum chaos conjecture associates the spectral statistics of a quantum system with abstract notions of quantum ergodicity. Such associations are taken to be of fundamental and sometimes defining importance for quantum chaos, but their practical relevance has been challenged by theoretical and experimental developments. Here, in counterpoint, we show that ergodic dynamics can be directly utilized for the preparation of quantum states with parametrically higher entanglement than generated by maximally scrambling dynamics such as in random unitary circuits. Our setting involves quantum systems coupled via a ``non-demolition'' interaction of conserved charges. We derive an exact relation between the evolving entanglement of an initial product state and a measure of spectral statistics of the interacting charges in this state. This connection is explained via a notion of Krylov vector ergodicity, tied to the ability of quantum dynamics to generate orthonormal states over time. We consider exploiting this phenomenon for the preparation of approximate Einstein-Podolsky-Rosen (EPR) states between complex systems, a crucial resource for tasks such as quantum teleportation. We quantitatively show that the transfer of operators between entangled systems, which underlies the utility of the EPR state, can be performed with parametrically larger capacity for entanglement generated via ergodic dynamics than with maximal scrambling. Our analysis suggests a direct application of ``ergodic'' spectral statistics as a potential resource for quantum information tasks.
\end{abstract}
%counterpoint is used here also in the sense of e.g. countermelody, a parallel thread that adds to the main melody, but apparently this is not an obvious reference.

\maketitle

\textit{Introduction---} \edit{The dynamics of complex quantum systems has turned out to be surprisingly resistant to a theoretical characterization in terms of sharp conceptual notions. Technical terms that have played foundational roles in discussing complex classical dynamics, including ``ergodicity'' and ``chaos'', have been applied to quantum systems in different (often fundamentally distinct~\footnote{For example, quantum ergodicity for Hamiltonian systems may refer to, e.g., (1) nondegeneracy and rational dependence properties of energy levels~\cite{vonNeumannThermalization, WuErgMix} that have no finite-time implications, (2) properties of time-averaged expectation values of observables~\cite{vonNeumannThermalization, PeresErgodic}, (3) the delocalization of energy eigenstates in convenient semiclassical bases~\cite{Shnirelman, CdV, ZelditchOG, ZelditchTransition, ZelditchMixing, Sunada, Zelditch, Anantharaman}, (4) as especially relevant for this work, the orthogonality properties of the observable-independent dynamics of states in the Hilbert space~\cite{dynamicalqergodicity, JTreconstruction} connected to the \textit{statistics} of energy levels~\cite{Mehta, Haake}.}) ways, but have fallen short of emulating their classical successes for the much richer phenomenology of quantum dynamics. This lack of precise notions in ``quantum chaos''~\cite{Haake}, as the subject is sometimes called for historical reasons~\cite{BerryChaology, BerryChaologyNotChaos, Berry227, PechukasWhatIsQuantumChaos, RudnickWhatIsQChaos}, has significantly limited the predictive power of our theories in this domain. Strikingly, no \textit{predictive} connections to chaos are known that apply universally to all systems \textit{without exceptions}. As a practical compromise, most developments rely on a \textit{statistical} approach to quantum dynamics: one tentatively forgoes the ambition of making precise predictions for individual systems, and instead comments on statistical correlations between different quantities that remain true for a wide variety of systems (which are then labeled ``typical'' systems~\cite{Haake, Mehta, DAlessio2016, KrylovReview}, essentially by definition).}
%A loose characterization is: come up with a statistical principle (based on universality properties of a measure like the Haar measure) and then assert that systems that behave like the Haar typical ones are typical (tautology?) and those that don't are not. We would like to identify more nontrivial connections [A slightly nontrivial thing e.g. in circuit models is we can show that smaller sub-ensembles can even behave close to Haar typical, but 2-designs are easy and spectral statistics is a 2-point thing, so is that really a surprise? Moreover the periodic orbit argument is quite general, just tells you that whatever averaging works for Haar typical also works for almost anything where the periodic orbits / closed Feynman paths are sufficiently statistically independent, which again feels like a statement about Haar designs.]
%The point here is that typical systems are defined by the Haar measure, and once one can compute something with the Haar measure (or even smaller ensembles), there is no longer any surprise in typicality, even if the calculation itself is very illuminating about the system. And in the interest of advancing our treatment of such questions, we are looking for more concrete things to say.

\edit{In the standard statistical scheme based on random matrix theory~\cite{Mehta, Haake, PeresErgodic}, one considers an ensemble of all possible quantum systems of a given type, and regards any system (or sub-ensemble) that behaves like a typical member of the larger ensemble by certain measures to be  ``quantum chaotic''. For example, in the case of Hamiltonian systems, the standards for complex dynamics are set by  random Hermitian matrices (acting as prototypes for the matrix elements of typical Hamiltonians in a generic basis). This approach has been adopted across a variety of fields,} ranging from many-body physics and statistical mechanics to quantum information and quantum gravity~\cite{ChaosComplexityRMT, KosProsen2018, ChanScrambling, ChalkerSum, ShenkerThouless, SSS, Suntajs2019}, \edit{encompassing a wide range of statistical trends in different measures that are held to probe ``chaos'' and ``complexity'' (mainly illustrating that many systems of sufficient complexity tend to exhibit different features similar to random matrices~\cite{ChaosComplexityRMT}). Nevertheless, a concrete prediction for quantum dynamics \textit{beyond randomness} remains conspicuously absent in this approach, which only serves to identify which systems are typical. This is quite different from classical ergodic theory~\cite{HalmosErgodic, Sinai1976, SinaiCornfeld}, whose dynamical content is far more substantial than statistical correlations in random ensembles of systems. For instance, ergodicity and chaos can be characterized in terms of observable independent notions such as indecomposability~\cite{HalmosErgodic, Sinai1976, SinaiCornfeld} and dynamical entropy~\cite{SinaiCornfeld}, and directly constrain the dynamics of observables~\cite{PlatoErgodic}.}

\edit{The purpose of this work is to illustrate that it is possible to refine the study of complex quantum dynamics beyond the statistical approach, by uncovering its direct operational relevance to tasks of interest in other domains.
%based on newly emerging connections between the different fields in which ``quantum chaos'' has been of key interest.
This entails a shift from viewing random matrix behavior as a statistical diagnosis of typicality to an operationally useful dynamical notion, with concrete predictions for specific tasks in individual systems. Our approach most directly focuses on the statistics of energy levels (or ``spectral statistics'' for short). Among the different statistical measures of ``quantum chaos'', spectral statistics has come to be widely regarded as being closest to \textit{the} defining measure~\cite{RudnickWhatIsQChaos, DAlessio2016, KosProsen2018}, in part due to being entirely observable-independent (as are the energy eigenvalues).
For similar reasons, the operational~\cite{QuantumOperationsReview} relevance of this ``defining'' measure has remained particularly elusive: no substantial connections to observables are known except for \textit{subleading} corrections to correlation functions~\cite{ChaosComplexityRMT, TorresHerreraSantos} that are challenging to even measure~\cite{SFFmeas, pSFF, pSFFexpt1, AdwaySantos}, much less connect to other tasks.}

\edit{\textit{Spectral statistics as a diagnostic---}} The study of spectral statistics has evolved two primary labels to qualitatively classify quantum dynamics. If the energy levels are arranged relatively regularly, with small spectral fluctuations around a regular integer lattice, then the label ``quantum chaotic'' is considered appropriate for the system~\cite{Haake}. Typically, such near-regular configurations are seen in the eigenvalue distributions of random matrices~\cite{Haake, Mehta}, \edit{motivating this label within the statistical approach} as codified in the quantum chaos conjecture~\cite{DKchaos, CGV, BerryStadium, BGS}, but also trivially in simple ergodic systems such as the harmonic oscillator \edit{(an ``atypical'' system that belongs outside this classification)}. In contrast, non-``quantum chaotic'' systems tend to show large spectral fluctuations, such as for a Poisson (random) distribution of points~\cite{BerryTabor} \edit{(but some of these systems can be classically ergodic and chaotic~\cite{BGGSarithmetic})}. \edit{A key shortcoming is that these labels have no rigorous connections to any intuitive notion of ``chaos'' or even \textit{other} measures of random matrix behavior in an \textit{individual} system (see Appendix~\ref{app:spectralstatistics} for details): e.g. a popular belief~\cite{DAlessio2016} connecting random matrix spectral statistics to the emergence of statistical mechanics and thermalization for \textit{local observables} in large systems is not well supported as a direct physical mechanism, despite co-occurring in typical systems. Specifically, rigorous operational criteria for quantum statistical mechanics can be formulated entirely in terms of finite-time correlation functions~\cite{dynamicalqthermalization, dynamicalpurestatethermalization, dynamicalqaperiodicity}, with no direct dependence on spectral statistics, and with forms of eigenstate statistics directly emerging as a consequence rather than an input. This is analogous to the classical observation~\cite{KhinchinStatMech, BricmontErgodicityCritique, GoldsteinErgodicityCritique, LebowitzAPS2021} that statistical mechanics is independent of formal ergodic dynamics. That the ``statistical definition of quantum chaos'' does not rigorously correspond to any familiar forms of ``chaos'' or statistical mechanics, in our view, makes this quantity most in need of a different approach.}
%We would therefore like to identify \textit{at least one} relevant physical process that is reliably predicted by spectral statistics, without exceptions.}
%Technically, spectral statistics is a set of measures, but this is a linguistic exercise: we can regard a complete set of quantities such as the SFF and phase of the trace, or even just the set of energy eigenvalues, as constituting our "measure".

\edit{\textit{Overview of results---} Our main result is that spectral statistics exhibits a \textit{direct correspondence} with entanglement generation in a specific quantum information transfer protocol, which constitutes a first prediction for an \textit{intuitive} and \textit{operationally useful} form of quantum dynamics unambiguously associated with random matrix spectral statistics. This is motivated by theoretical connections between ergodicity, random matrix statistics, and orthonormal state generation developed in previous work on quantum dynamics~\cite{dynamicalqergodicity} and 2D quantum gravity~\cite{JTreconstruction}.
%Our route to this prediction is based on the following chain of reasoning: spectral statistics has been identified with the presence of an observable-independent form of ergodic dynamics, corresponding to an exploration of orthonormal states in the Hilbert space~\cite{dynamicalqergodicity} (in analogy with exploring a classical phase space), with direct implications for wormhole dynamics in nonperturbative quantum gravity~\cite{JTreconstruction}. Wormholes have been regarded as ``dual'' to highly entangled quantum states (via a hypothesis dubbed ``ER = EPR''~\cite{ER_EPR}, an instance of the holographic duality~\cite{holography} between gravity and non-gravitational quantum systems), loosely suggesting that one look for a connection between spectral statistics and some form of entanglement. Here, we identify a concrete connection that is independent of gravity:
Specifically, we isolate a form of orthonormal state generation (``Krylov vector ergodicity''), demonstrably predicted by spectral statistics, that can be used to prepare states substantially close to maximally entangled states, namely Einstein-Podolsky-Rosen (EPR) states, through interactions that have been shown to be feasible to realize in experimental quantum platforms~\cite{QNDspin, SFFmeas}.}

\edit{EPR states are resource states for quantum information transfer, such as in quantum teleportation~\cite{NielsenChuang}. Despite a well-known ideal protocol for qubit systems~\cite{BennettTeleportation}, EPR state preparation in complex (e.g. high-dimensional) quantum systems~\cite{HighDimEntanglementReview, HighDimEntanglement1, HighDimEntanglement2} remains a significant challenge~\cite{TeleportationReview1, TeleportationReview2, KCRTeleportation} across a variety of experimental platforms, for which a generic solution is to use ``scrambling'' dynamics~\cite{HaydenPreskill, SekinoSusskind,LashkariFastScrambling, BentsenGuLucasScrambling, LucasEntanglementVsOTOC, dynamicalqspeedlimit, dynamicalqfastscrambling} (here, meaning generating high entanglement between subsystems, as operationally justified via information recovery protocols~\cite{YoshidaKitaev, YoshidaYao, scramblingSWAP}) to generate approximate EPR pairs.}

\edit{As an additional outcome of our approach, Krylov vector ergodicity can generate states that are parametrically closer to EPR states than scramblers for comparable system sizes, while being similarly ubiquitous (in ``typical'' systems) as a property of random matrices. From a practical standpoint,} the enhanced entanglement from ergodicity is accompanied by a significantly increased capacity of the entangled state to perform tasks such as quantum teleportation~\cite{BennettTeleportation}, compared to states generated by a typical scrambler
\edit{(requiring smaller Hilbert spaces for the same degree of entanglement or information transfer} [Eq.~\eqref{eq:operatorcapacityincrease}]\edit{)}. We highlight that our analysis \edit{allows} us to \edit{\textit{operationally}} separate the notions of ergodicity and scrambling (sometimes loosely associated with each other~\cite{ShenkerThouless, ChaosComplexityRMT} \edit{as typical ``signatures'' of ``quantum chaos''}), to respectively refer to orthonormal state generation in the Hilbert space~\cite{dynamicalqergodicity, JTreconstruction}, and strongly entangling dynamics~\cite{LashkariFastScrambling, BentsenGuLucasScrambling, LucasEntanglementVsOTOC, dynamicalqspeedlimit, dynamicalqfastscrambling}, \edit{each of which can be harnessed independently for teleportation via distinct protocols}. \edit{To draw a parallel with classical ergodic theory, our approach establishes spectral statistics as a precise observable-independent diagnostic (in analogy with indecomposability or dynamical entropy) of Krylov vector ergodicity and the dynamics of entanglement generation/information transfer (in analogy with time-averaged thermalization or chaos in all correlation functions, but arguably more ``operationally useful'').}
%The details of this approach are described below.
%

\textit{Information transfer---} Let us \edit{now} review the task at hand. For two systems with Hilbert spaces $\mathcal{H}_1$ and $\mathcal{H}_2$ each of dimension $d$, a maximally entangled EPR state~\cite{EPR, BennettTeleportation, NielsenChuang} must be of the form:
\begin{equation}
    \lvert \epr\rangle = \frac{1}{\sqrt{d}}\sum_{k=0}^{d-1} \lvert k\rangle_{1} \otimes \lvert k'\rangle_{2},
    \label{eq:EPRdef}
\end{equation}
where $\lvert k\rangle_1$ is an orthonormal basis for $\mathcal{H}_1$, as is $\lvert k'\rangle_2$ for $\mathcal{H}_2$. After preparing this state, one can move $\mathcal{H}_1$ and $\mathcal{H}_2$ to physically very different locations, and still transfer quantum information between the two locations via local operations and classical communication~\cite{NielsenChuang}, for example in quantum teleportation~\cite{BennettTeleportation, NielsenChuang, TeleportationReview1, TeleportationReview2}. This is due to the following crucial property of the EPR state, mathematically allowing the transfer of any operator $O_1$ acting on $\mathcal{H}_1$ to an equivalent operator $O_2$ acting on $\mathcal{H}_2$:
\begin{equation}
    O_1\lvert \epr\rangle = O_2^T\lvert \epr\rangle.
    \label{eq:eprtransfer}
\end{equation}
Here, ${}^T$ denotes a transpose of matrix elements, with $O_2$ being defined by the requirement that $O_1$ and $O_2$ have identical matrix elements in the respective $k$ and $k'$ bases: $\langle j\rvert O_1\lvert k\rangle_1 = \langle j'\rvert O_2\lvert k'\rangle_2 = \langle k'\rvert O_2^T\lvert j'\rangle_2$.

%So we didn't end up mentioning quantum operator stacks in the main text, but probably worth TeX-commenting on it anyway. The reason one would want to generate EPR states as opposed to using Hayden-Preskill type scrambling via entanglement in time (the A to C mechanism of \cite{scramblingSWAP})  for teleportation is that in the latter, once the scrambling is done, you can't insert new operators. But with an EPR state, we can keep inserting operators into one system until we finally project onto a pure state, and it will keep getting teleported. So this allows some delayed decisions. Moreover, operator insertion is "last in first out", like the stack in computer memory organization. The last operator to be inserted into A will become chronologically the first to act on B after transfer.

%It is theoretically straightforward to prepare an EPR state between (say) many qubits using Hadamard and CNOT gates~\cite{NielsenChuang}. However, a lack of sufficient control in complex experimental platforms necessitates more limited system-specific protocols~\cite{TeleportationReview1, TeleportationReview2, KCRTeleportation}, in which generating maximal entanglement is often challenging~\cite{HighDimEntanglementReview, HighDimEntanglement1, HighDimEntanglement2}. %This part is now stated above.
\edit{Where EPR states are difficult to generate,} a more generic alternative~\cite{YoshidaKitaev, YoshidaYao, scramblingSWAP} is to use some sufficiently complex ``scrambling'' interaction~\cite{HaydenPreskill} between a smaller system $\mathcal{H}_1$ and a larger system, whichever is available in a given platform, to generate highly entangled states. Any such entangled state can be written~\cite{SchrodingerHJSW, HJSW2, NielsenChuang} in the following form, where $\mathcal{H}_2$ is now some appropriate subspace of the larger system:
%This is strictly speaking just a variant of Schmidt decomposition, but S-HJW refs probably provide more historical motivation, and discuss the Schmidt form. Plus Schmidt's result is not quantum specific.
\begin{equation}
    \lvert \Psi\rangle = \frac{1}{\sqrt{d}}\sum_{k=0}^{d-1} \lvert k\rangle_1 \otimes \lvert \psi_k\rangle_2.
    \label{eq:genericentangledstates}
\end{equation}
Here, $\lvert k\rangle_1$ remains an orthonormal basis for $\mathcal{H}_1$, but the $\lvert \psi_k\rangle_2$ are no longer necessarily orthonormal. Clearly, the closer the $\lvert \psi_k\rangle_2$ are to forming an orthonormal basis for $\mathcal{H}_2$, the more accurately the highly entangled state can emulate the information transfer property \eqref{eq:eprtransfer} of an EPR state.

From this viewpoint, our main result is that ergodic dynamics, where it can be harnessed, can generate entangled states \eqref{eq:genericentangledstates} in which the $\lvert \psi_k\rangle_2$ are much closer to being an orthonormal basis for $\mathcal{H}_2$ than generic scrambling dynamics in many-body systems. In particular, for a family of ``highly ergodic'' initial states, our protocol generates demonstrably more entanglement than even infinite temperature ``maximal'' scramblers such as random quantum circuits; for more generic initial states, the advantage in entanglement is limited to being over finite temperature scramblers such as many complex local Hamiltonians.

\textit{Krylov vector ergodicity---} Our \edit{specific} setting \edit{to harness spectral statistics} involves a system~\cite{NielsenChuang} Alice with a Hilbert space $\mathcal{H}_A$ of dimension $d_A = d$ (replacing $\mathcal{H}_1$ above), which contains the information to be encoded. Another system Bob with a $(d_B \geq d_A)$-dimensional Hilbert space $\mathcal{H}_B$, is one to which we would like to transfer this information (in the subspace $\mathcal{H}_2$). By assumption, Alice's system is simple and allows the application of all desired quantum operations, such as (but not necessarily) a quantum computer. Bob's system is complex, and the best we can do is apply a certain Hamiltonian $H_B$. Specifically, we couple the two systems by a non-demolition interaction \edit{similar to time-estimation settings~\cite{TimeEstimationPageWootters, TimeEstimation1, TimeEstimation2}} (see App.~\ref{app:KrylovErgodicity} for generalizations):
\begin{equation}
    H = N_A \otimes H_B,
    \label{eq:Hint_QND_1}
\end{equation}
%Maybe the best way to realize something like this with some sufficiently complex $H_B$ is if $N_A$ is (up to shifting and scaling) the $S_z$ spin operator for a $d_A$-level spin, and $\mathcal{B}_z = H_B$ is a magnetic field in the z-direction, generated by some sufficiently complex combination of charged currents/quasiparticles/other constituents in the subsystem $B$ that it has random matrix statistics, for example. This is very similar to the setting in \cite{QNDspin}, but there $H_B$ is just a randomly weighted some over spins so it is more likely to generically have Poisson statistics.
in which $N_A$ is some ``number'' operator in $A$ with eigenvalues $\lbrace 0, 1, \ldots, d_A-1\rbrace$ and eigenstates $\lvert n\rangle_A$. Interactions of this type can be engineered, for example, in quantum dots~\cite{QNDspin} and Rydberg atoms~\cite{SFFmeas}. We will take $A$ to be initialized to an equal superposition of $N_A$-eigenstates, as is conventional in such protocols:
\begin{equation}
    \lvert \chi\rangle_A = \frac{1}{\sqrt{d_A}}\sum_{n=0}^{d_A-1} \lvert n\rangle_A.
\end{equation}
We will also assume that $B$ is initialized to an unspecified initial state $\lvert \phi\rangle_B$ (owing to Bob's limited control), whose properties determine certain advantages of our protocol.

The state of the combined system at time $t_0$ is
\begin{equation}
    \lvert \Psi(t_0)\rangle = e^{-i H t_0} \lvert \chi\rangle_A \otimes \lvert \phi\rangle_B.
    \label{eq:generalstatedynamics}
\end{equation}
Defining the unitary operator $U_{t_0} = \exp(-iH_B t_0)$, we see that this time-dependent state can be expressed as:
\begin{equation}
    \lvert \Psi(t_0)\rangle = \frac{1}{\sqrt{d_A}} \sum_{n=0}^{d_A-1} \lvert n\rangle_A \otimes U_{t_0}^n\lvert \phi\rangle_B.
    \label{eq:ergodic_entangled_state}
\end{equation}
From the arguments pertaining to Eq.~\eqref{eq:genericentangledstates}, it is clear that the closeness of $\lvert \Psi(t_0)\rangle$ to an EPR state is decided by the orthonormality of the $d_A$-element set of \textit{Krylov} vectors~\cite{KrylovOG, Sinai1976, KrylovBook} generated by successive actions of $U_{t_0}$ on $\lvert \phi\rangle_B$:
\begin{equation}
    \mathcal{K}_{d_A}(U_{t_0}, \lvert \phi\rangle_B) = \left\lbrace \lvert \phi\rangle_B, U_{t_0}\lvert \phi\rangle_B, \ldots, U_{t_0}^{d_A-1}\lvert \phi\rangle_B \right\rbrace.
    \label{eq:Krylovset}
\end{equation}
These vectors track~\cite{Sinai1976} the dynamics of $\lvert \phi\rangle_B$, and span the subspace $\mathcal{H}_2 \subseteq \mathcal{H}_B$ in which Bob can encode information from Alice.

At a conceptual level, the orthonormality of the Krylov set is connected to spectral statistics as follows. In Ref.~\cite{dynamicalqergodicity}, it is shown (adapted to our present language) that the smallness of spectral fluctuations, such as for random matrices, allows the \textit{existence} of an orthonormal basis in $\mathcal{H}_B$ such that a $d_B$-element Krylov set of each basis element has a strong overlap with the original basis, which provides a quantum notion of dynamical ergodicity (in the sense of dynamics exploring a complete orthonormal basis). While it is difficult to make a direct quantitative connection of our present protocol with this notion, the conceptual mechanism at play here is similar: we will see that suppressed spectral fluctuations, both of the energy levels and of the initial state $\lvert \phi\rangle_B$, lead to an increased orthonormality of the Krylov set.
%But note that $P_S(t) \geq K(t)$ \cite{dynamicalqspeedlimit} does allow some vague relation, in the sense that $\sum_k p_{\phi_k}(t) \geq K(t)$ when $\lvert \phi_k\rangle$ is an orthonormal basis of pure states. But then, $K(t)$ only has an indirect relation with cyclic ergodicity through the variance of mode fluctuations, but cyclic ergodicity also cares about the distribution of mode fluctuations.

We call the latter phenomenon ``Krylov vector ergodicity''. As an aside, we note its potential to bridge the notions of dynamical ergodicity~\cite{dynamicalqergodicity} and Krylov state complexity~\cite{spreadcomplexity, KrylovReview} (delocalization in a 
Krylov set related to actions of $H_B$ rather than $U_{t_0}$), whose correspondence has been conjectured in Ref.~\cite{JTreconstruction} based on theoretical quantum gravity observations\edit{, as an extension of the ``complexity = volume'' conjecture~\cite{SusskindCV, ComplexityAnything}}. \edit{Qualitatively, this correspondence is due to the fact that both ergodicity and Krylov complexity are different measures of orthogonality in quantum dynamics; however, Krylov vector ergodicity provides an experimentally accessible measure of this orthogonality via the protocol proposed here, whereas Krylov state complexity~\cite{KrylovReview} is not known to be directly accessible.}
%The other interesting question is does this have any implications for ER = EPR, because \cite{JTreconstruction} connects ergodicity to wormhole dynamics (ER), while this paper connects ergodicity to EPR state generation. And of course, an ``ergodic theory'' approach, if not directly dynamical ergodicity, is connected to fast scrambling~\cite{dynamicalqspeedlimit, dynamicalqfastscrambling}. What other close connections might there be between quantum information, quantum dynamics, and quantum gravity in this more direct sense? How many of them is ergodicity likely to be related to in some way?

\textit{Ergodic entanglement generation---} We will measure entanglement between $A$ and $B$, and therefore the orthonormality of the Krylov set in Eq.~\eqref{eq:Krylovset} \edit{i.e. Krylov vector ergodicity}, by means of the purity~\cite{NielsenChuang} (whose negative logarithm gives the $2$nd R\'{e}nyi entanglement entropy~\cite{HorodeckiEntanglementReview, AndreasPurityRM}), as justified in App.~\ref{app:KrylovErgodicity}:
\begin{equation}
    \mathcal{P}(t) = \Tr_B\left[\left\lbrace \vphantom{\sum} \Tr_A\left(\lvert \Psi(t)\rangle_{AB}\langle\Psi(t)\rvert\right)\right\rbrace^2\right].
    \label{eq:puritydef}
\end{equation}
In all cases, we have $d_A^{-1} \leq \mathcal{P}(t) \leq 1$, with the lower limit $d_A^{-1}$ corresponding to maximal entanglement as for EPR states. For the state in Eq.~\eqref{eq:ergodic_entangled_state}, it follows that
\begin{equation}
    \mathcal{P}(t_0) = \frac{1}{d_A} + \frac{2}{d_A}\sum_{\tau = 1}^{d_A-1}\left(1-\frac{\tau}{d_A}\right) p_{\phi}(\tau t_0),
    \label{eq:purity_and_probability_deja_vu}
\end{equation}
where the deviation from maximal entanglement is directly given by the \textit{dynamics} of the return probability $p_{\phi}(t)$ of the initial state:
\begin{align}
    p_{\phi}(t) &\equiv \left\lvert\langle \phi\rvert e^{-iH_Bt}\lvert \phi\rangle_B\right\rvert^2 \nonumber \\
                &= \sum_{n,m=0}^{d_B-1} \lvert \phi(E_n)\rvert^2 \lvert \phi(E_m)\rvert^2 e^{-i(E_n-E_m)t}.
                \label{eq:returnprobabilitydef}
\end{align}
Here, $\lvert E_n\rangle_B$ refers to the energy eigenstates of $H_B$ with respective eigenvalues $E_n$, and $\phi(E_n) = \langle E_n\vert \phi\rangle_B$ are the components of the initial state in this basis. The essential intuition behind Eq.~\eqref{eq:purity_and_probability_deja_vu} is that a larger overlap with the initial state over time hinders the ability of dynamics to effectively explore the rest of the Hilbert space to generate more orthonormal states~\cite{dynamicalqergodicity}, therefore increasing purity and reducing entanglement. \edit{Moreover, Krylov vector ergodicity can be ``promoted'' to a state-independent property of a system (as is spectral statistics), e.g. for a given parameter set $(t_0, d_A)$, by minimizing $\mathcal{P}(t_0)$ over all initial states $\lvert \phi\rangle_B$ in $B$.}
%Carrying out this minimization for an orthonormal basis instead of a state might replace the return probability with the SFF, as in \cite{dynamicalqspeedlimit, dynamicalqfastscrambling}.

%\comm{Update terms: RMT $E_n$, Poisson $E_n$, uniform $\phi$, random $\phi$?} Or maybe doesn't really add much.
The quantity $p_{\phi}(t)$ is sensitive to both the fluctuations of the energy levels $E_n$, and the fluctuations of the state components $\phi(E_n)$, in a manner that has been studied in several works, e.g.~\cite{WilkieBrumerReturnProbabilities, TorresHerreraSantos}. The details of this behavior are
%reviewed
\edit{illustrated in a concrete class of examples} in App.~\ref{app:theexcruciatingcalcdetails}, \edit{with more general comments in App.~\ref{app:probfluctuations}}; here, we note the relevant qualitative trends. %\edit{We note that the expressions below completely capture the leading order behavior of the relevant quantities, as supported by a specific illustration for random matrix ensembles in Appendix~\ref{app:theexcruciatingcalcdetails}.}
After a time $t > t_{\ramp}$, determined by the energy scale of fluctuations, we can write $p_{\phi}(t)$ as a sum of contributions from spectral fluctuations and initial state fluctuations:
\begin{equation}
    p_{\phi}(t > t_{\ramp}) \sim f_E(t) + f_{\phi}(t).
    \label{eq:rampandnoise}
\end{equation}
The contribution from spectral statistics is, for $t$ not too long:
\begin{equation}
    f_E(t) \sim \begin{dcases}
        O(t/d_B^2),&\ \text{ergodic $E_n$}, \\
        1/d_B,&\ \text{Poisson $E_n$},
    \end{dcases}
    \label{eq:fE}
\end{equation}
reflecting the smaller spectral fluctuations expected for ergodic systems (typically, with random matrix statistics). The contribution from the initial state depends on whether $\phi(E)$ varies smoothly as a function of energy $E_n$. This is the case for special initial states, such as discrete Fourier transforms of the energy eigenstates. %or coherent Gibbs states~\cite{coherentGibbs}.
More generic initial states in complex systems, however, fluctuate randomly. We have:
\begin{equation}
    f_{\phi}(t) \sim \begin{dcases}
        0,&\ \text{smooth } \phi(E), \\
        1/d_B,&\ \text{generic } \phi(E).
    \end{dcases}
    \label{eq:fphi}
\end{equation}
The combination of random matrix statistics (or smaller spectral fluctuations) and a smooth $\phi(E)$ corresponds to a highly ergodic Krylov set.
From Eq.~\eqref{eq:purity_and_probability_deja_vu}, we have at the (not too long) time $t_0 > t_{\ramp}$ (with ergodic $E_n$ applying for $d_A \ll d_B$, and effectively becoming Poisson for $d_A \gg d_B$):
\begin{equation}
    \mathcal{P}(t_0) = \frac{1}{d_A}+\begin{dcases}
        O\left(d_A/d_B^2\right),&\ \text{ergodic $E_n$ \& smooth $\phi$}, \\
        1/d_B,&\ \left\lbrace\substack{\text{ergodic $E_n$ \& generic $\phi$}, \\ \text{Poisson $E_n$ \& smooth $\phi$},}\right. \\
        2 /d_B,&\ \text{Poisson $E_n$ \& generic $\phi$},
    \end{dcases}
    \label{eq:purity_summary1}
\end{equation}

It is worth comparing these results to entanglement generation via scrambling dynamics. Under direct time evolution (outside our protocol), a typical ``scrambling'' Hamiltonian $H_s$ with local interactions between $A$ and $B$ generally thermalizes initial states to a state-dependent finite temperature $\beta^{-1}$ at all sufficiently long times $t > t_s$ (up to quantum recurrences~\cite{QuantumRecurrences, BrownSusskind2}), incapable of generating maximal entanglement. Quantitatively, Ref.~\cite[Eq. (28)]{lu2019renyi} suggests that for generic finite temperature local Hamiltonian dynamics,
%one may have
%For a Gaussian density of states, which is believed to be generic
\begin{equation}
    \mathcal{P}_{\beta}(t > t_s) \sim \max\lbrace d_A^{-1+c\beta^2}, d_B^{-1+c\beta^2}\rbrace,
    \label{eq:ThermalPurity}
\end{equation}
where $c > 0$ is a nonvanishing $O(1)$ parameter that depends on $d_A$ and $d_B$, which is not maximal entanglement even to leading order for $\beta > 0$.
\edit{In particular, the volume-law coefficient of the R\'{e}nyi entropy~\cite{HorodeckiEntanglementReview}, $S_2 = - \log_2 \mathcal{P} \sim f_V \max\lbrace N_A, N_B\rbrace$, is given by $f_V \propto 1-c\beta^2$  in a many-body system of $N_A$ qubits.} 
%$f_V = \ln 2$ for a maximal volume law coefficient, which corresponds to $c\beta^2 = 0$, but usually we have $f_V < \ln 2$ for finite-temperature dynamics.
Infinite temperature scrambling to $\beta = 0$, such as for strongly nonlocal Hamiltonians~\cite{LiuXiaoBalents_SYKEntanglement, ThermalNEEs, dynamicalqspeedlimit} or random quantum circuits~\cite{RandomCircuitEntanglement}, can generate maximal entanglement up to subleading corrections:
\begin{equation}
    \mathcal{P}_{\beta=0}(t > t_s) = \frac{1}{d_A}+\frac{1}{d_B},
    \label{eq:HaarPurity}
\end{equation}
as estimated for \textit{ideal} random systems~\cite{PageSubsystem, GorinSeligman2001, VinayakZnidaric, RandomCircuitEntanglement, pSFF} (neglecting higher order terms in $d_A$ and $d_B$, including subleading spectral corrections~\cite{GorinSeligman2001, VinayakZnidaric, TorresHerreraSantos}). Larger deviations from minimal purity may occur for physical nonlocal Hamiltonians~\cite{ThermalNEEs, LiuXiaoBalents_SYKEntanglement}.
From Eq.~\eqref{eq:purity_summary1}, even with a generic random wavefunction $\phi(E)$ and a (possibly nonscrambling) Hamiltonian $H_B$, a random matrix spectrum produces at least as much entanglement in our protocol as an infinite-temperature scrambler in Eq.~\eqref{eq:HaarPurity}, with considerably better performance for a family of ``smooth'' initial states that experience highly ergodic dynamics.

\textit{Operator transfer capacity---} Having identified an enhancement in entanglement generation from ergodic dynamics, it is worth asking to what extent this enhancement actually matters in quantum information processing. We will find a notable advantage for operator transfer as in Eq.~\eqref{eq:eprtransfer}, implying corresponding gains in tasks such as quantum teleportation (relative to scrambling dynamics).

To understand these gains, it is useful to write our entangled state as the EPR state with a \textit{pre-loaded} operator:
\begin{equation}
    \lvert \Psi(t)\rangle = R_B(t)\lvert \epr\rangle,
\end{equation}
where $R_B$ is a \textit{general} linear transformation in $\mathcal{H}_B$, transforming the basis $\lvert k'\rangle_2$ in Eq.~\eqref{eq:EPRdef} to the states $\lvert \psi_k\rangle_2$ in Eq.~\eqref{eq:genericentangledstates}, such as the Krylov set in Eq. \eqref{eq:Krylovset}. This transformation determines the purity via $\mathcal{P}(t) = \Tr_B[(R_B(t) R_B^\dagger(t))^2]/d_A^2$.
%Note that $R_B$ is more precisely called $R_2$, as it annihilates all states in $\mathcal{H}_B$ that are orthogonal to R_2
We can now identify a generalized operator transfer property:
\begin{equation}
    O_A \lvert \Psi(t)\rangle = R_B(t) O_B^T\lvert \epr\rangle,
\end{equation}
where $O_A$ in Alice's system, acting on our entangled state, behaves as if it is mapped to $R_B(t) O_B^T$ in Bob's system by an ideal EPR state. For faithful transfer, we demand that the overlaps, given by trace inner products, of any two operators $O_A$, $P_A$ with Alice are approximately preserved \edit{(i.e., the transfer is nearly isometric, see also \cite{scramblingSWAP})}:
\begin{equation}
     \Tr_B\left[\left\lbrace R_B(t) O_B^T\right\rbrace \left\lbrace R_B(t) P_B^T \right\rbrace^\dagger \right]\simeq \Tr_A\left[O_A P_A^\dagger\right].
     \label{eq:isometricencoding_approx}
\end{equation}
%The formal motivation for this criterion will be described in upcoming work on encoding maps.
Operationally, the overlaps $d_A^{-2}\lvert \Tr_A[O_A P_A^\dagger]\rvert^2$ constrain~\cite{dynamicalqspeedlimit} the similarity of sets of states in $A$ under the actions of the two operators $\lvert \langle O^\dagger\psi\vert P^\dagger \psi\rangle_A\rvert^2$ (e.g., $O_A^\dagger$ and $P_A^\dagger$ may create excitations when acting on a Fock state $\lvert\psi\rangle_A$). This allows quantum teleportation~\cite{BennettTeleportation, NielsenChuang, TeleportationReview1, TeleportationReview2} via a $\langle \Psi(t)\rvert$-projective measurement of $B$ with a third system $A'$ formally identical to $A$, teleporting any $P_{A'}^\dagger$ in $A'$ to ${P_A^{\text{out} \dagger}} \equiv [R_A^T(t) R_A^{T\dagger}(t)] P_{A}^\dagger \simeq P_A^\dagger$ in $A$.

%related ideas will be discussed in upcoming work~\cite{scramblingSWAP}.
%\comm{Update isometry discussion}
%The reason this has to look so contrived with $P_A^\dagger$ instead of $P_A$ is because we have $O_A$ acting on $\lvert \Psi(t)\rangle$ or the EPR state. If we had called that $O_A^\dagger$ instead, this would be much simpler in terms of notation: because A is the output system and O_A is the probe, in some sense. Maybe this does remain thematically consistent with prior notation in the literature if one thinks of $P_A^\dagger$ as a creation operator acting on some reference vacuum state $\lvert 0\rangle_A$ before teleporting the resultant excitations.

In App.~\ref{app:operatortransfer}, we formulate this criterion more precisely to allow for a relative error $0 \leq \epsilon < 1$ in this inner product. For operator transfer with an error of at most $\epsilon$, we find that the purity must satisfy:
\begin{equation}
    \mathcal{P}(t) \leq \frac{1}{d_A}+\frac{\epsilon^2}{\kappa d_A^\gamma},
    \label{eq:purityscramblingcriterion1}
\end{equation}
With $\kappa = 1$, for $\gamma = 1$, this gives a strict necessary condition for the successful transfer of \textit{all} operators~\footnote{The maximal scrambling criterion used in \cite{dynamicalqspeedlimit, dynamicalqfastscrambling} is in turn a necessary condition for this to hold with $\epsilon = o(1)$.}, which we also expect to be close to sufficient $(\gamma \approx 1)$ in generic cases;
%rationale: Let the initial state be chosen from a basis $\lbrace \lvert k\rangle\rbrace$. The density operator in $A$ at time t is $\rho_A(t)$, let its traceless part be $\Delta rho(t) = \rho(t)-\idop/d_A$. Then, $\mathcal{P}(t) \leq (1+\epsilon^2)/d_A$ is equivalent to $\sum_{k,j} \lvert \Delta \rho_{kj}\rvert^2 < \epsilon^2/d_A$. This implies that for *each* $k$, $\rho_{kk}(t) = (1+\epsilon^2)/d_A$, which corresponds to the return probability of the initial state for the choice of k giving the initial state, and the other overlaps called $Q_{kj}$ in \cite{dynamicalqspeedlimit}. For $\epsilon = o(1)$, this implies $P_{kj}(t) = 1/d_A + o(1/d_A)$, which corresponds to the maximal scrambling condition in those Refs on identfying $d_A$ with $d_S$. But it is perhaps not sufficient, because of the purity-probability inequality in \cite{dynamicalqfastscrambling} (one can imagine that the off-diagonals of the density operator are potentially large).
%Implication: in addition to determining the validity of equilibrium statistical mechanics, our previous scrambling condition also sets a bound on the time after which \textit{sustained} operator transfer is possible between the coupled systems. But  this is implicitly a tripartite problem: in those Refs, the equivalent of $B$ (namely $E$) starts out in a mixed state that must be purified with an auxiliary system R (for Hawking Radiation). Operator transfer then becomes possible from the equivalent of $A$ (namely $S$) to the combined system of $E$ and $R$.
for $\gamma = 2$ we get a rigorous sufficient condition. With $\gamma = 0$, we get a necessary and sufficient condition for the transfer of \textit{typical}~\cite{tumulka_CT, CanonicalTypicalityPSW} (i.e. almost all) operators $O_A$ in $A$ when $\kappa \gg 1$ is a large $O(1)$ constant.
%Note that this $\gamma = 0$ case means that for a high-dimensional Hilbert space of $A$, virtually having even some $\Theta(1)$ purity is sufficient for teleporting typical operators with resolution $\epsilon = \Theta(1)$: we do not need anywhere near maximal entanglement for typical operators unless our resolution is sensitive to the dimension of $\mathcal{H}_A$.
\edit{Interestingly, this illustrates that spectral statistics measures the isometric encoding of ``time'' into (ergodic) dynamics~\cite{dynamicalqergodicity, JTreconstruction}, whereas scrambling performs the instantaneous isometric encoding of smaller subsystems into larger ones~\cite{scramblingSWAP}, both of which can be independently harnessed via $\mathcal{P}(t)$ for teleportation.}

Let us now use Eq.~\eqref{eq:purityscramblingcriterion1} to contrast ergodic entanglement generation with scramblers. By Eq.~\eqref{eq:ThermalPurity}, typical local Hamiltonian scramblers do not generate sufficient entanglement for transferring \textit{all} operators as their purity saturates at $\gamma = 1-c\beta^2 < 1$. But a comparable Hamiltonian realized in Bob's system via Eq.~\eqref{eq:Hint_QND_1} is capable of complete operator transfer due to ergodicity, as are infinite temperature $\beta = 0$ scramblers. For these systems, the subleading terms in Eqs.~\eqref{eq:purity_summary1} and \eqref{eq:HaarPurity} set a direct lower limit on the dimension $d_B$ of Bob's system required to faithfully encode information from Alice, where we regard a smaller lower limit on $d_B$ as a diagnostic of increased operator transfer capacity:
\begin{equation}
    d_B \gtrsim \begin{dcases}
        \left(\kappa d_A^{1+\gamma}/\epsilon^2\right)^{1/2},&\ \text{ergodic $E_n$ \& smooth $\phi$}, \\
        \varphi \kappa d_A^\gamma /\epsilon^2,&\ \text{generic $\phi$, OR $\beta = 0$}.
    \end{dcases}
    \label{eq:operatorcapacityincrease}
\end{equation}
%the $\gtrsim$ is because we ignore any small constants in the $O(d_A/d_B^2)$ term
In the second line, $\varphi = 2$ for Poisson statistics and $\varphi = 1$ otherwise.
%In the second line, $\varphi = 1$ for a random matrix spectrum or ideal $\beta = 0$ scrambler, while $\varphi = 2$ for Poisson statistics.
Further specifics depend on $\gamma$ and how $\epsilon$ scales with $d_A$.
As an illustrative case with potential practical relevance, let us consider operator transfer for $\epsilon = \delta d_A^{-1/2}$ with $\delta \ll 1$. Such a resolution is necessary for teleportation to preserve overlaps between typical random states $\lvert\langle \psi_1\vert \psi_2\rangle_A\rvert \sim d_A^{-1/2}$ in $\mathcal{H}_A$, as required to probe e.g. thermalization~\cite{tumulka_CT, CanonicalTypicalityPSW} in $A$.
Here, with highly ergodic dynamics (an ergodic spectrum \textit{and} a smooth initial state), successful teleportation for \textit{all} operators is guaranteed ($\gamma = 2$) with $d_B \gtrsim d_A^2/\delta$ but may generically occur ($\gamma \approx 1$) even with $d_B \gg d_A^{3/2}/\delta$, and for \textit{typical} operators ($\gamma = 0$) with only $d_B \gg d_A/\delta$. For $\beta = 0$ scramblers, we can teleport \textit{all} operators only with ($\gamma = 1$) at least $d_B \gtrsim d_A^2/\delta^2$, and \textit{typical} operators ($\gamma = 0$) with $d_B \gg d_A/\delta^2$. Even with generic initial states, the ``ergodic'' teleportation of \textit{typical} operators has comparable capacity to $\beta = 0$ scramblers, and retains a factor of $2$ advantage over Poisson statistics. For $\beta > 0$ scramblers, achieving such a small error is impossible with any $d_B$. These cases illustrate a clear power-law advantage (in $\delta$ or $d_A$, especially for $\gamma \approx 1$) of ergodicity in operator transfer capacity (via the smallness of Bob's system).
%, which carries over into protocols that rely on the EPR state such as quantum teleportation.
%Analyzing cases in this way is more conclusive and definite without having to evaluate the higher order purities for ergodic entanglement generation (to find the distribution of the $r_k$), which remains an interesting task for future work.

\textit{Discussion---} We have shown that ergodicity can be harnessed, even in the standard context of generating EPR states, to generate parametrically higher entanglement than scrambling dynamics for tasks such as quantum teleportation. This suggests a practical motivation for studying the spectral statistics of large-but-not-thermodynamically-large systems, to realize a simple goal of maximizing operator transfer capacity in highly entangled states. We regard this possibility as surprising to the extent that physical applications of even \textit{classical} ergodic dynamics are scarce, outside formal theoretical insights~\cite{PlatoErgodic}. \edit{In Ref.~\cite{dynamicalqaperiodicity}, we further show that a ``partial'' form of Krylov vector ergodicity, in a regime independent of spectral statistics, provides operational criteria for finite-time macroscopic thermalization.}

%Most importantly, our approach emphasizes a shift in viewpoint, from considering spectral statistics as a qualitative ``defining'' notion for ``quantum chaos'', whose implications remain rather vague, to a resource for quantum information processing with a demonstrable theoretical advantage. It is conceivable that other operational applications of spectral statistics exist, perhaps more significant than outlined here, that deserve to be explored; see also \cite{QuantumOperationsReview} for the significance of operational notions in quantum information.
%shift in viewpoint now moved to intro

In conclusion, it is important to emphasize a qualifier. Our approach should be regarded as a theoretical proof-of-principle for the practical utility of spectral statistics, but its potential for \textit{complete} experimental realization remains to be seen. The primary challenge is the complexity of preparing a ``smooth'' initial state that shows maximally ergodic dynamics, corresponding to the top row of Eq.~\eqref{eq:fphi}.
Nevertheless, even without the ability to easily prepare such a state, which diminishes its advantage over (ideal) infinite temperature scramblers, our protocol continues to have the potential for present-day use cases. A notable feature is its ability to generate enhanced entanglement comparable to infinite temperature scramblers even with generic initial states in an ergodic system, while continuing to use e.g. a local finite-temperature scrambling Hamiltonian
%(or even a non-ideal $\beta = 0$ Hamiltonian)
for Bob's system in Eq.~\eqref{eq:Hint_QND_1} that may otherwise be incapable of generating such entanglement.

\begin{acknowledgments}
   We thank Andrew Lucas for useful discussions. We acknowledge related collaborations with Chris Akers, Edwin Chaparro, Muhammad Miskeen Khan, Andrew Lucas, and Ana Maria Rey. This work was supported by the Heising-Simons Foundation under Grant 2024-4848.
\end{acknowledgments}

\begin{appendix}

\setcounter{secnumdepth}{4}   %needs to be put in by hand for the prl template, but the pra template does not need this. However, for some reason, the two look different, and the prl template is somewhat easier to read.
\setcounter{section}{0}

\section{What does spectral statistics tell us?}
\label{app:spectralstatistics}

\edit{Now we will elaborate on the lack of a rigorous connection between random matrix spectral statistics, sometimes taken to define ``quantum chaos''~\cite{DAlessio2016, KosProsen2018}, and other observable phenomena usually associated with chaos and thermalization. This necessitates our search in this paper for a physically interesting phenomenon that is \textit{undeniably} associated with random matrix spectral statistics.}

\edit{Historically, the motivation for studying spectral statistics stems from ``quantum chaology'': the study of quantum properties of classically chaotic systems~\cite{BerryChaology, BerryChaologyNotChaos, Berry227, Haake, Anantharaman}. While many quantized chaotic systems show random matrix spectral statistics~\cite{DKchaos, CGV, BerryStadium, BGS}, the same was found to be true for various non-chaotic, ergodic systems~\cite{GiraudWDwithoutPO, Lozej2021, CasatiWang}. On the other hand, the most standard examples of ``maximally'' chaotic and ergodic classical systems: Bernoulli shifts~\cite{BV, Saraceno, bakeranomalieserg}, the cat map~\cite{keating1991cat}, and systems with arithmetic chaos~\cite{BGGSarithmetic, BraunHaakearithmetic}, are all strong exceptions to this trend that do not have random matrix spectra in their standard quantizations (the latter even being nearly Poisson). In fact, the presence of random matrix statistics in these cases depends strongly on the chosen quantization procedure rather than the presence of ergodicity or chaos~\cite{Saraceno, bakeranomalieserg, keating1991cat, BraunHaakearithmetic}, and may in this sense be attributed to the ``statistical typicality'' properties of the quantization procedure much more than the dynamical properties of the classical system.}
%The statistical typicality depends on how periodic orbits are captured by the quantization, but periodic orbits are a measure zero set in classical dynamics and do not correspond to standard mesasure theoretic dynamical notions.

%\edit{A key issue is that despite the association of spectral statistics with ``quantum chaos'', no concrete relation has been discovered between the former and any intuitive notion that reminds one of ``chaos''.}

%\comm{(also dubbed~\cite{Haake} ``universal'' in a statistical sense, but with several exceptions)}

\edit{Fully quantum many-body systems generally do not have a classical limit, and one can hope for more clarity within the quantum regime without any dependence on a quantization procedure. Here}, it is \edit{traditionally~\cite{PeresErgodic}} believed that random matrix spectral statistics is responsible  for the validity of quantum statistical mechanics, via \edit{e.g.} the eigenstate thermalization hypothesis~\cite{vonNeumannThermalization, JensenShankarETH, deutsch1991eth, srednicki1994eth, srednicki1999eth, rigol2008eth, DAlessio2016, deutsch2018eth} (which amounts to an assumption on the statistics of energy \textit{eigenstates} rather than eigenvalues, \edit{which are mathematically independent of each other}). %This motivation~\cite{DAlessio2016}, though not quite firmly established, particularly underlies the study of spectral statistics  in the form of diagnosing ``many-body quantum chaos''~\cite{DAlessio2016,ChaosComplexityRMT, KosProsen2018, ChanScrambling, ChalkerSum, ShenkerThouless, SSS, Suntajs2019}.
However, we believe that this traditional viewpoint is (by now) significantly challenged by theoretical and experimental developments. On the experimental side, recent efforts~\cite{SFFmeas, pSFF, pSFFexpt1, AdwaySantos} affirm that spectral statistics is most accessible in smaller systems of $\sim 10$ particles, rather than the thermodynamically large systems of e.g. $\sim 10^{23}$ particles relevant for statistical mechanics; the potential for improved thermodynamic scaling appears to be bleak. On the theoretical side, it turns out to be possible to formulate quantum statistical mechanics in a form with equivalent and even stronger implications for experimentally accessible observables and timescales than eigenstate thermalization~\cite{dynamicalqthermalization, dynamicalpurestatethermalization, dynamicalqaperiodicity}, \edit{without making any assumptions on eigenstate statistics that instead follow as an output from finite-time correlators, and with no requirements on the spectra}.
%Moreover, eigenstate thermalization is easier to establish via operational statistical mechanics than spectral statistics, so even if the point of spectral statistics is to diagnose eigenstate thermalization (which we do not claim that it does) it is harder to establish than the thing it is suggested to be a diagnostic of.
\edit{Incidentally, for quantized classical systems, some of these eigenstate properties can be connected quite directly to classical thermalization dynamics~\cite{Shnirelman, CdV, ZelditchOG, ZelditchTransition, ZelditchMixing, Sunada, Zelditch, Anantharaman, dynamicalqthermalization, dynamicalqaperiodicity} without specific sensitivity to the quantization procedure, unlike spectral statistics, further highlighting their stark distinction.}

To us, it therefore appears fairly compelling that random matrix spectral statistics is not the \textit{direct} mechanism behind statistical mechanics, though they often co-occur due to the complexity \edit{(or typicality)} of the associated quantum systems (which should not be taken as evidence of a causal link, as supported by explicit counterexamples \edit{involving local systems}~\cite{JensenShankarETH, MaganWu}). Nevertheless, spectral statistics remains a widely available resource that can potentially be harnessed: generic complex quantum systems have nontrivial spectral statistics resembling those of random matrices~\cite{Haake, Mehta}. This compels us to explore if this ubiquitous resource can be put to work for some specific task in systems of, say, a few $10$s of particles as can be manipulated in present-day quantum experiments, rather than the more macroscopic considerations of statistical mechanics for which it may well remain inaccessible. \edit{As one concrete instance, we propose the quantum information protocol considered in this work}.

\section{Multidimensional Krylov vector ergodicity}
\label{app:KrylovErgodicity}

The most general nondemolition interaction between $A$ and $B$, of which Eq.~\eqref{eq:Hint_QND_1} is a special case, is a Hamiltonian that couples conserved charges between the two systems:
\begin{equation}
    H = H_A \otimes \idop_B + \idop_A\otimes H_B + \sum_{k} q_{Ak} \otimes Q_{Bk}.
    \label{eq:Hint_QND_gen}
\end{equation}
Here, $H_A$ and $H_B$ represent the independent dynamics of $A$ and $B$, while $q_{Ak}$ and $Q_{Bk}$ are mutually commuting conserved charges: $[q_{Ak},H_A] = [q_{Ak},q_{Aj}] = 0$, and $[Q_{Bk}, H_B] = [Q_{Bk},Q_{Bj}] = 0$.
%It is sufficient for these charges to only be approximately conserved over the timescale of interest.
Eq.~\eqref{eq:Hint_QND_gen} may be easier to implement in a wider range of systems~\cite{QNDspin, QNDspinQDOT} than Eq.~\eqref{eq:Hint_QND_1}.

Let $\lvert n\rangle_A$ and $\lvert E_m\rangle_B$ be joint eigenbases of these operators in $A$ and $B$, with $q_{Ak}$ eigenvalues $q_{Ak,n}$ and $Q_{Bk}$ eigenvalues $Q_{Bk,m}$. Then for the same dynamics as Eq.~\eqref{eq:generalstatedynamics}, we obtain
\begin{equation}
    \lvert \Psi(t_0)\rangle = \frac{e^{-i H_0 t_0}}{\sqrt{d_A}}\sum_{n=0}^{d_A-1}\lvert n\rangle_A\ \otimes \prod_k e^{-i(q_{Ak,n} t_0) Q_{Bk}} \lvert \phi\rangle_B,
\end{equation}
where $H_0 = H_A \otimes \idop_B + \idop_A\otimes H_B$. The relevant $d_A$-element Krylov set is now determined by the multidimensional group of symmetry transformations $U_{t_0,k} = \exp(-i Q_{Bk} t_0)$ generated by the different conserved charges, over $d_A$ different sets of multidimensional ``angles'' $q_{Ak,n}$ determining the duration of each generating transformation:
\begin{equation}
    \mathcal{K}_{q_{Aj}}\left(U_{t_0,k},\lvert\phi\rangle_B\right) = \left\lbrace \lvert \mathcal{K}_n\rangle =\prod_k U_{t_0,k}^{q_{Ak,n}}\lvert \phi\rangle_B,\ \forall n \in \mathbb{Z}_{d_A}\right\rbrace.
    \label{eq:multidimensionalKrylov}
\end{equation}
The orthonormality of this basis is sensitive to the spectral statistics of each charge $Q_{Bk}$. For sufficiently complex charges, the $Q_{Bk}$ can have nontrivial spectral statistics. For the setting of Refs.~\cite{QNDspin, QNDspinQDOT} where $B$ is a quantum dot, the charge $Q_{B1}$ is a randomly weighted sum of local spins, and may generically have Poisson statistics. If any $Q_{Bk}$ corresponds to a sufficiently complex operator in $B$, then it may have random matrix statistics.

Recall that the Krylov set spans a subspace $\mathcal{H}_2 \subseteq \mathcal{H}_B$ of dimension $d_2 \leq d_A$, as we have $d_A$ Krylov vectors. We now extend $\mathcal{H}_2$ by introducing auxiliary dimensions if necessary to make $d_2 = d_A$. If $\lvert j\rangle_2$ is an orthonormal basis for this extended $\mathcal{H}_2$ space, we can write the Krylov set as a general linear transformation of this orthonormal basis:
\begin{equation}
    \lvert \mathcal{K}_j\rangle_2 = R_2\lvert j\rangle_2.
\end{equation}
In the ideal case where the Krylov set is orthonormal, $R_2$ is a unitary operator satisfying $R_2 R_2^\dagger = R_2^\dagger R_2 =\idop_2$. More generally, we can measure ``Krylov vector ergodicity'' via the closeness of $R_2$ to a unitary transformation, e.g. the smallness of
\begin{equation}
   \eta_2 \equiv \frac{1}{d_2}\Tr_2\left[(R_2 R_2^\dagger)^2\right]-1,
   \label{eq:KrylovErgodicityMeasure}
\end{equation}
with $\eta_2 = 0$ for a perfect unitary transformation (which generates ideal EPR states) and $\eta_2 > 0$ otherwise. This is because $R_2 R_2^\dagger$ and $R_2^\dagger R_2$ are positive operators (with nonnegative eigenvalues) and $\Tr_2[R_2 R_2^\dagger] = d_2$ is fixed by $\langle \mathcal{K}_j\vert \mathcal{K}_j\rangle = 1$, which follows from Eq.~\eqref{eq:multidimensionalKrylov}; therefore, $\eta_2 = 0$ implies that $R_2R_2^\dagger = R_2^\dagger R_2 = \idop_2$. We emphasize that, unlike dynamical ergodicity in Ref.~\cite{dynamicalqergodicity} which is either present or absent in a system (which is useful to make quantitative connections to random matrix statistics), it is more convenient to regard Krylov vector ergodicity as a ``continuous'' measure whose degree (of absence) is given by $\eta_2$. An explicit calculation of $\eta_2$ verifies its connection to multidimensional return probabilities of $\lvert \phi\rangle_B$ under the symmetry transformations $U_{t_0,k}$, generalizing Eq.~\eqref{eq:purity_and_probability_deja_vu}.

Finally, if we evaluate the purity $\mathcal{P}(t_0)$ of the state $\lvert \Psi(t_0)\rangle$ according to Eq.~\eqref{eq:puritydef}, we get
\begin{equation}
    \mathcal{P}(t_0) = \frac{1}{d_A^2}\Tr_2\left[(R_2 R_2^\dagger)^2\right] = \frac{1+\eta_2}{d_A},
    \label{eq:purity_and_ergodicity}
\end{equation}
verifying that the deviation from minimal purity $1/d_A$ (or maximal entanglement) directly measures Krylov vector ergodicity in the above sense. Such expressions generalize to scrambling dynamics if one writes $\rho_A(t) = R_2^T R_2^{T\dagger}$, where $\rho_A(t_0) = \Tr_B[\lvert \Psi(t_0)\rangle\langle\Psi(t_0)\rvert]$ is the (positive) reduced density operator of the entangled state in $A$.
%Any positive operator can be factorized into (non-unique) square roots, so one choice is $R_2^T = [\rho_A(t)]^{1/2}$
More generally, we can work with ``higher purities'' (related to the $\alpha$-R\'{e}nyi entropy~\cite{HorodeckiEntanglementReview}, with $\alpha = 2$ giving the conventional purity),
\begin{equation}
    \mathcal{P}_\alpha(t_0) \equiv \Tr_A\left[\rho^{\alpha}_A(t_0)\right] = \frac{1}{d_A^{\alpha}}\Tr_2\left[(R_2 R_2^\dagger)^{\alpha}\right],
    \label{eq:higherpurities}
\end{equation}
which are given by e.g. more complicated functions of the amplitudes $\langle \phi\rvert U_{t_0}^{\tau_j}\lvert \phi\rangle_B$ than $p_{\phi}(t)$, involving multiple times $\tau_j$ with a cyclicity condition $\sum_j \tau_j = 0$ and some associated phase interference effects for the case of Eq.~\eqref{eq:Hint_QND_1}; we will therefore leave a detailed study of these quantities (for $\alpha \neq 2$) for future work.

%There may also be something funny about mutual spectral form factors $|\Tr(e^{-iqQt})}^2$...is this supposed to be the spectral statistics of $q$ or $Q$ or both? Are there consistency conditions here that could suggest something?

\section{Illustration in ideal statistical ensembles}
\label{app:theexcruciatingcalcdetails}

\edit{Here, we will illustrate our claims for ideal statistical ensembles modeled by random matrices, which (within the statistical approach~\cite{PeresErgodic, DAlessio2016}) capture the behavior of ``typical'' systems. In particular, we will contrast operator transfer for entanglement generated via the ergodicity of a random Hamiltonian $H_B$ in $B$ coupled to $A$ via Eq.~\eqref{eq:Hint_QND_1} as opposed to the scrambling dynamics of another random Hamiltonian $H_{AB}$ directly acting on both systems. More general considerations are discussed in App.~\ref{app:probfluctuations}.}

\edit{Formally, we will consider $4$ statistical ensembles for the initial state and $H_B$, all of which can be described by a single expression with Boolean parameters $c_{\phi}, c_E \in \lbrace 0,1\rbrace$: (1) initial states $\lvert \phi\rangle_B$ that are (a) averaged over Haar random~\cite{Mehta, Haake} states ($c_\phi = 1$) or (b) unbiased ($c_\phi = 0$) with $\lvert\phi\rangle_B = d^{-1/2}\sum_n E_n$, with (2) averaging over spectral statistics~\cite{Haake, Mehta} that follow (a) the Circular Unitary Ensemble (CUE) ($c_E = 1$) or (b) Poisson ($c_E = 0$) statistics. Then, we can write the ensemble-averaged $p_{\phi}(t)$ as~\cite{WilkieBrumerReturnProbabilities} (cf. Eq.~\eqref{eq:rampandnoise}):
\begin{equation}
    \overline{p_{\phi}(t)} = \underbrace{\frac{c_\phi}{d_B+1}}_{f_\phi(t)} + \underbrace{\frac{d_B}{d_B+1} \overline{K(t)}}_{f_E(t)},
    \label{eq:RMT_p}
\end{equation}
where $K(t)$ is the spectral form factor~\cite{Haake, BerrySpectralRigidity}, whose ensemble average for CUE or Poisson statistics with a uniform (ensemble-averaged) density of states $E \in [0,2\pi)$ at discrete times $t \in \mathbb{Z}$ is~\cite{Haake, Mehta}:
\begin{equation}
    \overline{K(t \in \mathbb{Z})} \equiv \frac{1}{d_B^2}\overline{\sum_{n,m}e^{i(E_n-E_m)t}} = \delta_{t,0}+\frac{1}{d_B}\left[1-c_E\left(1-\frac{|t|}{d_B}\right)_+\right].
    \label{eq:RMT_SFF}
\end{equation}
Here, $(x)_+ = \max(x,0)$. The first term in Eq.~\eqref{eq:RMT_p} is identified with $f_{\phi}(t)$, and the second term (given by Eq.~\eqref{eq:RMT_SFF}) for $\lvert t\rvert \ll d_B$ with $f_E(t)$ (here, behaving as $t/d_B^2$ for CUE and $1$ for Poisson), thereby showing the leading order behaviors in Eqs.~\eqref{eq:fE} and \eqref{eq:fphi}. We note that for CUE, $t_{\ramp} = 1$ in $\mathbb{Z}$, but this can differ for other systems~\cite{dynamicalqergodicity, ShenkerThouless}. Additionally, these are ensemble-averaged expressions, but those for typical systems fluctuate erratically around these values~\cite{PrangeSFF}, motivating the order-of-magnitude expressions in Eqs.~\eqref{eq:fE} and \eqref{eq:fphi} (see also App.~\ref{app:probfluctuations}).
}

\edit{For the purity $\mathcal{P}(t_0)$, we substitute Eqs.~\eqref{eq:RMT_p} and \eqref{eq:RMT_SFF} in Eq.~\eqref{eq:purity_and_probability_deja_vu} with $t_0 \in \mathbb{Z}$; the sum over times $\tau t_0$ tames the fluctuations between systems, allowing us to use ensemble-averaged expressions even for typical individual systems in the ensemble~\cite{KosProsen2018}. We get
\begin{equation}
    \mathcal{P}(t_0) = \frac{1}{d_A}+\frac{c_\phi(d_A-1)}{d_A(d_B+1)}+\frac{d_A^{1+c_E}-1}{(1+2c_E)d_A d_B^{c_E}(d_B+1)}.
\end{equation}
To leading order in $d_A,d_B \gg 1$, this gives the expressions in Eq.~\eqref{eq:purity_summary1}, with $c_{\phi} = 0,1$ respectively corresponding to smooth (unbiased) and generic (Haar) states, and $c_E = 0,1$ to Poisson and ergodic (CUE) $E_n$.
}

%\section{Operator transfer capacity}
%\label{app:operatortransfercapacity}

\section{Spectral fluctuations in return probabilities}
\label{app:probfluctuations}

Now, we will review some quantitative details of the behavior of $p_{\phi}(t)$ in Eq.~\eqref{eq:returnprobabilitydef}, and some miscellaneous issues.
As there are much better resources discussing this behavior with more scope than this short Appendix, we will primarily just point to the literature here with brief accompanying details.

%\textit{Review---}
At early times, $p_{\phi}(t)$ decays smoothly from $1$ at $t=0$, in a way that depends on the coarse-grained profile of $\phi(E)$. When $p_{\phi}(t) \sim O(1/d_B)$ after a time $t > t_{f}$, it fluctuates erratically~\cite{PrangeSFF} depending on the $E_n$ and $\phi(E_n)$. This is usually not analytically tractable, but reasonable estimates have been made in the literature~\cite{WilkieBrumerReturnProbabilities, TorresHerreraSantos, ThoulessRelaxation} assuming that these fluctuations are sufficiently random. In most systems, to a good approximation, these are often independent, and add up independently as in Eq.~\eqref{eq:rampandnoise}. As Eq.~\eqref{eq:purity_and_probability_deja_vu} sums over $p_{\phi}(t)$ at different times, it is reasonable to approximate these (non-negative) fluctuations by their average value, as we have done above.

The spectral contribution $f_E(t)$ may be isolated (up to fluctuations) by setting $\phi(E_n) = 1/d_B$, for which $f_{\phi}(t) = 0$; for this choice, $p_{\phi}(t)$ exactly corresponds to the state-independent ``spectral form factor'', whose phenomenology has been explored in detail in e.g.~\cite{Haake, Mehta, ShenkerThouless, ChaosComplexityRMT, refRampPlateau2, KosProsen2018, ChanScrambling, ChalkerSum, SSS, Suntajs2019, SFFmeas, pSFF, pSFFexpt1, dynamicalqergodicity, dynamicalqspeedlimit, dynamicalqfastscrambling}. After a system-specific~\cite{ShenkerThouless} time $t_{\ramp} > t_f$ (called the ``ramp'' or ``Thouless'' time), which may be~\cite{BertiniProsen, BKPRMTSFFCMP} as small as $O(1)$ and smaller or larger than the scrambling time $t_s$, $f_E(t) \sim t/d_B^2$ (the ``ramp'') for a random matrix spectrum up to the Heisenberg time $t_{\htime}\sim 2\pi \dos(E)$ determined by the typically $O(d_B)$ density of states $\dos(E)$. Our ``ergodic $E_n$'' expression in Eq.~\eqref{eq:fE} assumes $t_0 \lll t_{\htime} \sim d_B$ in the sense that $(\log t_0)/(\log d_B) \approx 0$ [which also restricts $t_{\ramp} < t_0$], and $t_0 d_A \ll t_{\htime}$ (therefore 
$d_A \ll d_B$), quantifying which times we regard as ``not too long''. Poisson statistics corresponds to a constant $f_E(t) \sim 1/d_B$ shortly after $t_{f}$, as does random matrix statistics after $t \sim t_{\htime}$. For simplicity, we neglect other cases of intermediate statistics.
%$t_0 \approx O(1)$: this can be a soft $O(1)$, including e.g. log factors.

%\comm{Complement ramp time with fluctuations time $t_f$?}

The wavefunction contribution $f_{\phi}(t)$ has been considered in e.g. \cite{WilkieBrumerReturnProbabilities, TorresHerreraSantos, ThoulessRelaxation}, and is generally smaller for more delocalized wavefunctions in $E$. Eq.~\eqref{eq:fphi} corresponds to fully delocalized wavefunctions with maximal randomness, as for a mathematically (Haar) random state, in the second row. \textit{Any} reduction in randomness would decrease this offset. The first row with $f_{\phi}(t) \sim 0$ includes $\phi(E_n) = 1/d_B$, but also other smooth wavepackets such as Gaussians $\phi(E) \sim \exp(-(E-E_0)^2/4\sigma^2)$ or coherent Gibbs states~\cite{coherentGibbs} $\phi(E) \sim \exp(-\beta E/2)$ at $E = E_n$; the absence of large fluctuations implies negligible $f_{\phi}(t)$ a while after $t_{f}$.

%\textit{Misc. details---}
To generate maximal entanglement, we would ideally like the range of times $[t_0, t_0d_A]$ to be as close as possible to $t_{\ramp}$ for minimal $f_{E}(t)$. A spectrum or $\phi(E)$ with sharp features can obscure the ramp time due to slowly decaying oscillations~\cite{ShenkerThouless, dynamicalqspeedlimit}, so a smooth wavepacket is also our best bet for the earliest detectable $t_{\ramp}$. Discrete-time systems with a bounded spectrum such as quantum circuits are generically subject to such oscillations~\cite{dynamicalqspeedlimit} unless one can fine-tune $t_0$ and $N_A$, so a continuous time Hamiltonian may tolerate errors in $t_0$ or $N_A$ better.

%To avoid state preparation issues, we note that if we initialize $d_B$ to a maximally mixed state $\rho_B = \idop_B/d_B$ (which is easily prepared by maximal scrambling with a larger system $\mathcal{H}_R$ that may be ignored after state preparation~\cite{HaydenPreskill}) instead of a pure state $\lvert \phi\rangle_B$, then $p_{\phi}(t)$ in Eq.~\eqref{eq:purity_and_ergodicity} is replaced by the exact spectral form factor $K(t) = \lvert\Tr_B(e^{-i H_B t})/d_B\rvert^2$. The resulting purity is $\mathcal{P}(t_0) = 1/d_A+O(d_A/d_B^2)$; however, an infinite temperature scrambler generates $\mathcal{P}(t > t_s) = 1/d_A + O(1/d_B^2)$, which is more efficient. Nevertheless, this scenario may still be useful to \textit{formally} generate near-minimal purity in $A$ by leveraging the dynamical ergodicity of finite temperature e.g. local Hamiltonians $H_B$ in Eq.~\eqref{eq:Hint_QND_1} compared to Eq.~\eqref{eq:ThermalPurity}, without state preparation difficulties, but also without any operator transfer capacity from $A$ to $B$ due to the mixed initial state.

% \comm{A key benefit of using an ergodic Hamiltonian, as opposed to a quantum circuit, is as follows. Quantum circuits have to be fine tuned to early times for having small entanglement, even for preparing EPR states. But after $t_0 > t_{\ramp}$, we can basically work with any time and do not need to fine tune the time for generating $O(1/d_A)$ entanglement. This feels like it could be important to emphasize maybe even in the main text, if there's space?}

\section{Details of operator transfer capacity}
\label{app:operatortransfer}

Here, we quantify the criterion for faithful operator encoding in Eq.~\eqref{eq:isometricencoding_approx}. By the Cauchy-Schwarz inequality~\cite{ByronFuller},
\begin{equation}
    \left\lvert \Tr_A[O_A P_A^\dagger]\right\vert \leq \sqrt{\Tr_A[O_A O_A^\dagger] \Tr_A[P_A P_A^\dagger]}.
    \label{eq:CauchySchwarz}
\end{equation}
We will let the right hand side set the scale relative to which we measure the error $\epsilon$ in the inner products of encoded operators $O_A \to R_2 O_2^T$, in the notation of App.~\ref{app:KrylovErgodicity}. We consider the entangled state $\lvert \Psi(t_0)\rangle$ to be able to successfully encode operators with an error of $\epsilon$ or less if $\Delta_2(O_A,P_A) \leq \epsilon$, where
\begin{equation}
    \Delta_2(O_A,P_A)\equiv\frac{\left\lvert \Tr_2\left[\left\lbrace R_2 O_2^T\right\rbrace \left\lbrace R_2 P_2^T \right\rbrace^\dagger \right] - \Tr_A[O_A P_A^\dagger]\right\rvert}{\sqrt{\Tr_A[O_A O_A^\dagger] \Tr_A[P_A P_A^\dagger]}}.
\end{equation}
As $\mathcal{H}_2$ is isomorphic to $\mathcal{H}_A$ by construction, it follows that (with some implicit transposes and cycling of operators):
%which leave the trace invariant
\begin{equation}
    \Delta_2(O_A,P_A) = \frac{\left\lvert \Tr_2\left[(R_2^\dagger R_2-\idop_2) O_2^TP_2^{T\dagger}\right]\right\rvert}{\sqrt{\Tr_A[O_A O_A^\dagger] \Tr_A[P_A P_A^\dagger]}}.
\end{equation}
For quantum teleportation~\cite{NielsenChuang} to $A$ from a system $\mathcal{H}_{A'}$ isomorphic to $\mathcal{H}_A$, this allows the teleported operator $P_A^{\text{out}\dagger} = (R_2^T R_2^{T\dagger}) P_{A}^\dagger$ to have the same overlap with other operators $O_A$ acting on $A$ as would $P_{A}^\dagger$, up to the error $\epsilon$. We take $\epsilon < 1$ as a requirement for better performance than $P_A^{\text{out}} = 0$.

Let us first set $P_A = O_A$. Then, $(O_2^\dagger O_2)^T$ is a positive operator with nonnegative eigenvalues $\omega_j \geq 0$ and orthonormal eigenstates $\lvert \omega_j\rangle$. The above expression becomes:
\begin{equation}
    \Delta_2(O_A,O_A) = \frac{\left\lvert\sum_{j} \omega_j \langle \omega_j \rvert (R_2^\dagger R_2 -\idop_2)\lvert \omega_j\rangle \right\rvert}{\sum_j \omega_j}.
\end{equation}
Consequently, a necessary and sufficient condition for $\Delta_2(O_A,O_A) \leq \epsilon$ for any $O_A$ is that
\begin{equation}
    \max_{\lvert \psi\rangle \in \mathcal{H}_2 :\ \langle \psi\vert \psi\rangle = 1}\ \left\lvert \langle\psi\rvert (R_2^\dagger R_2-\idop_2)\lvert \psi\rangle\right\rvert \leq \epsilon.
    \label{eq:R2eigenvalueconstraint}
\end{equation}
This measure can, in principle, be accessed through the higher purities $\mathcal{P}_\alpha(t_0)$ in Eq.~\eqref{eq:higherpurities}, for example $\lim_{\alpha\to\infty}\mathcal{P}_{2\alpha}^{1/(2\alpha)}(t_0)$. However, as the behavior of these quantities is complicated to evaluate, we will attempt to relate Eq.~\eqref{eq:R2eigenvalueconstraint} to $\mathcal{P}(t_0)$.
Let $r_k \geq 0$ denote the eigenvalues of $R_2^\dagger R_2$, whose mean is $1$ as $\Tr_2[R_2 R_2^\dagger]/d_2 = 1$; the above condition enforces ${\lvert r_k-1\rvert} \leq \epsilon$. Then, $\eta_2 = d_2^{-1}\sum_k(r_k-1)^2$ measures the variance of the $r_k$ around $1$, and the Bhatia-Davis inequality~\cite{BhatiaDavis} gives
\begin{equation}
    \eta_2 \leq (r_{\max}-1)(1-r_{\min}) \implies \eta_2 \leq \epsilon^2,
    \label{eq:eta2constraint}
\end{equation}
where $r_{\max}$ and $r_{\min}$ are the maximum and minimum of the $r_k$, as a necessary condition for Eq.~\eqref{eq:R2eigenvalueconstraint} to be valid, implying Eq.~\eqref{eq:purityscramblingcriterion1} with $\gamma = 1$, $\kappa = 1$. For a sufficient condition based on $\eta_2$, the best we can do is impose that $\eta_2 \leq \epsilon^2/d_2$, corresponding to Eq.~\eqref{eq:purityscramblingcriterion1} with $\gamma = 2$ and $\kappa = 1$. Further details depend on the distribution of the $r_k$, and may be interesting to explore in future work e.g. via $\mathcal{P}_{\alpha}(t_0)$. But we expect that $\gamma \approx 1$ with $\kappa = 1$ is generically sufficient; e.g. if the $r_k$ follow a Gaussian distribution, then we expect $(r_{\max}-1)$ and $(1-r_{\min})$ to be~\cite{GaussianMaxdist1, GaussianMaxdist2, GaussianMaxdist3} near $\sqrt{2 \eta_2 \ln d_2}$, corresponding to Eq.~\eqref{eq:purityscramblingcriterion1} with $\gamma = 1$ and $\kappa \sim \ln d_A$ (which can be absorbed into $\gamma \approx 1$ with $\kappa = 1$).
%The Gaussian distribution may be partly motivated by noting that the higher order purities involving higher order loops of Krylov actions may primarily be dominated by Wick's theorem type terms due to the randomness of phases, such as motivated by Berry's diagonal approximation for the SFF, which should be suggestive of a Gaussian. But the cyclic permutation symmetry of periodic orbits may obviously introduce some corrections.

%Typo in prev version, corrected now.

For $O_A \neq P_A$, we can write $(R_2^\dagger R_2-\idop_2) = M_2^2$, where $M_2$ is non-Hermitian with eigenvalues $\sqrt{r_k-1}$ and eigenvectors $\lvert r_k\rangle$. Then, the Cauchy-Schwarz inequality~\eqref{eq:CauchySchwarz} with $O_A \to M_2 O_2^T$ and $P_A^\dagger \to P_2^{T\dagger} M_2$ gives (noting that $M_2 M_2^\dagger = M_2^\dagger M_2$)
\begin{equation}
    \Delta_2(O_A,P_A) \leq \sqrt{\Delta_2^M(O_A,O_A) \Delta_2^M(P_A,P_A)},
\end{equation}
%So this is what Cauchy-Schwarz looks like for an indefinite metric: the indefinite metric gets replaced by its definite variant for the factors within the square root.
where $\Delta_2^{M}$ corresponds to $\Delta_2$ with $(R_2^\dagger R_2-\idop_2)$ replaced by $M_2^\dagger M_2$, whose eigenvalues are $\lvert r_k-1\rvert$. Thus, $\lvert r_k-1\rvert \leq \epsilon$ is also necessary and sufficient for $\Delta_2(O_A,P_A) \leq \epsilon$ as with Eq.~\eqref{eq:R2eigenvalueconstraint}, implying the same criteria as above in terms of $\eta_2$.

We can also obtain a distribution-independent constraint for the transfer of ``typical'' operators in $\mathcal{H}_A$, i.e., $O_A = V_A \overline{O}_A V_A^\dagger$ and $P_A = V_A \overline{P}_A V_A^\dagger$, where $\overline{O}_A$, $\overline{P}_A$ are arbitrary operators and $V_A$ is a (Haar) random unitary~\cite{Mehta, Haake} in $\mathcal{H}_A$ (but $P_A$ is not random relative to $O_A$). To leading order in the Hilbert space dimension $d$, the effect of averaging two-point correlators over Haar random unitaries is to ``decouple'' (or ``thermalize'') the correlator~\cite{CotlerHunterJones2}, e.g., for $\Tr[\Lambda] = 0$,
%This doesn't have to be a Haar average, just a sufficiently random one
%The idea is that $O_A$ is random relative to the eigenbasis $\lvert r_k\rangle$.
\begin{equation}
    \left\langle \left\lvert \Tr[V \Omega V^\dagger \Lambda]\right\rvert^2\right\rangle_V \simeq \frac{1}{d^2}\Tr[\Omega^\dagger \Omega]\Tr[\Lambda^\dagger\Lambda].
\end{equation}
%Full expression: $\left\langle \left\lvert \Tr[V \Omega V^\dagger \Lambda]\right\rvert^2\right\rangle_V \simeq \frac{1}{d^2}\left(\left\lvert \Tr[\Omega]\right\rvert^2\left\lvert \Tr[\Lambda]\right\rvert^2 + \Tr[\Omega^\dagger \Omega]\Tr[\Lambda^\dagger\Lambda]\right)$.
%We can get this just by a Wick's theorem approximation for the complex random variables that are the matrix elements of $V_A$ because we're at leading order and we don't need the full Haar average: we just use \langle V_{ij} V_{kl}^\ast\rangle \simeq (1/d)\delta_{ik} \delta_{jl}$ and $\langle V_{ij} V_{kl}\rangle \simeq 0$ to leading order.
Applying this relation to $|\Delta_2(O_A, P_A)|^2$ with $\Omega_A = (\overline{P}_A^\dagger \overline{O}_A)^T$ and $\Lambda = (R_2^\dagger R_2 -\idop_2)$, we get for large $d_A$
%We take $P_A$ to also have the same randomness as $O_A$
\begin{equation}
    \left\langle \left\lvert \Delta_2(O_A, P_A)\right\rvert^2\right\rangle_{V_A} \simeq \frac{\eta_2}{d_A} \frac{\Tr_A[\Omega_A \Omega_A^\dagger]}{\Tr_A[O_A O_A^\dagger] \Tr_A[P_A P_A^\dagger]}.
    \label{eq:typicaloperatoroverlapfidelity}
\end{equation}
For a large fraction $f$ of individual choices of $V_A$, $|\Delta_2(O_A,P_A)\rvert^2$ may fluctuate~\cite{PrangeSFF} up to as much as a large constant factor $\kappa \gg 1$ times this Haar average (with $f\to 1$ as $\kappa \to \infty$, even after taking $d_A \to\infty$). To have an error of at most $\epsilon$ for such ``typical'' operators, noting that $\Tr_A[\Omega_A \Omega_A^\dagger] \leq \Tr_A[O_A O_A^\dagger] \Tr_A[P_A P_A^\dagger]$, it is necessary and sufficient that $\eta_2 \leq \epsilon^2 d_A/\kappa$, leading to the condition for transferring ``typical'' or ``almost all'' operators in Eq.~\eqref{eq:purityscramblingcriterion1}.

\end{appendix}

\bibliography{TangledBibliography}

@article{vonNeumannThermalization,
  title={Proof of the ergodic theorem and the {H-theorem} in quantum mechanics. {Translation of: Beweis des Ergodensatzes und des {H-Theorems} in der neuen Mechanik}},
  author={von Neumann, John},
  journal={Eur. Phys. J. H},
  volume={35},
  number={2},
  pages={201--237},
  year={2010},
  publisher={Springer},
url={https://doi.org/10.1140/epjh/e2010-00008-5},
note={{Original (in German)}: Zeit. f\"{u}r Phys. 57, 30 (1929)}
}

@article{tumulka_CT,
  title={Canonical typicality},
  author={Goldstein, Sheldon and Lebowitz, Joel L and Tumulka, Roderich and Zangh{\`{i}}, Nino},
  journal={Phys. Rev. Lett.},
  volume={96},
  number={5},
  pages={050403},
  year={2006},
  publisher={APS},
  url={https://doi.org/10.1103/PhysRevLett.96.050403}
}

@article{CanonicalTypicalityPSW,
author = {Popescu, Sandu and Short, Anthony J and Winter, Andreas},
doi = {10.1038/nphys444},
journal = {Nat. Phys.},
number = {11},
pages = {754--758},
title = {Entanglement and the foundations of statistical mechanics},
url = {https://doi.org/10.1038/nphys444},
volume = {2},
year = {2006}
}

@article{deutsch1991eth,
  title = {Quantum statistical mechanics in a closed system},
  author = {Deutsch, J. M.},
  journal = {Phys. Rev. A},
  volume = {43},
  issue = {4},
  pages = {2046--2049},
  numpages = {0},
  year = {1991},
  publisher = {American Physical Society},
  doi = {10.1103/PhysRevA.43.2046},
  url = {https://link.aps.org/doi/10.1103/PhysRevA.43.2046}
}

@article{srednicki1994eth,
  title = {Chaos and quantum thermalization},
  author = {Srednicki, Mark},
  journal = {Phys. Rev. E},
  volume = {50},
  issue = {2},
  pages = {888--901},
  numpages = {0},
  year = {1994},
  publisher = {American Physical Society},
  doi = {10.1103/PhysRevE.50.888},
  url = {https://link.aps.org/doi/10.1103/PhysRevE.50.888}
}

@article{srednicki1999eth,
author = {Srednicki, Mark},
doi = {10.1088/0305-4470/32/7/007},
journal = {J. Phys. A: Math. Gen.},
number = {7},
pages = {1163--1175},
publisher = {IOP Publishing},
title = {{The approach to thermal equilibrium in quantized chaotic systems}},
url = {http://dx.doi.org/10.1088/0305-4470/32/7/007},
volume = {32},
year = {1999}
}

@article{rigol2008eth,
author = {Rigol, Marcos and Dunjko, Vanja and Olshanii, Maxim},
doi = {10.1038/nature06838},
journal = {Nature},
number = {7189},
pages = {854--858},
title = {{Thermalization and its mechanism for generic isolated quantum systems}},
url = {https://doi.org/10.1038/nature06838},
volume = {452},
year = {2008}
}

@article{deutsch2018eth,
author = {Deutsch, Joshua M},
doi = {10.1088/1361-6633/aac9f1},
journal = {Rep. Prog. Phys.},
number = {8},
pages = {082001},
publisher = {IOP Publishing},
title = {{Eigenstate thermalization hypothesis}},
url = {http://dx.doi.org/10.1088/1361-6633/aac9f1},
volume = {81},
year = {2018}
}

@article{DAlessio2016,
author = {D'Alessio, Luca and Kafri, Yariv and Polkovnikov, Anatoli and Rigol, Marcos},
doi = {10.1080/00018732.2016.1198134},
journal = {Adv. Phys.},
number = {3},
pages = {239--362},
title = {{From quantum chaos and eigenstate thermalization to statistical mechanics and thermodynamics}},
url = {https://doi.org/10.1080/00018732.2016.1198134},
volume = {65},
year = {2016}
}

@book{Mehta,
  title={Random matrices},
  author={Mehta, Madan Lal},
  year={2004},
  publisher={Elsevier},
  isbn={978-0-12-088409-4}
}

@book{Haake,
title={Quantum signatures of chaos},
author={Haake, Fritz},
year={2001},
publisher={Springer, Berlin, Heidelberg},
url={https://doi.org/10.1007/978-3-662-04506-0}}

@article{PrangeSFF,
  title={The spectral form factor is not self-averaging},
  author={Prange, R E},
  journal={Phys. Rev. Lett.},
  volume={78},
  number={12},
  pages={2280},
  year={1997},
  publisher={APS},
  url={https://doi.org/10.1103/PhysRevLett.78.2280}
}

@article{ChalkerSum,
  title={Spectral statistics and many-body quantum chaos with conserved charge},
  author={Friedman, Aaron J and Chan, Amos and De Luca, Andrea and Chalker, JT},
  journal={Phys. Rev. Lett.},
  volume={123},
  number={21},
  pages={210603},
  year={2019},
  publisher={APS},
  url={https://doi.org/10.1103/PhysRevLett.123.210603}
}

@article{SFFmeas,
  title={Monitoring quantum simulators via quantum nondemolition couplings to atomic clock qubits},
  author={Vasilyev, Denis V and Grankin, Andrey and Baranov, Mikhail A and Sieberer, Lukas M and Zoller, Peter},
  journal={PRX Quantum},
  volume={1},
  number={2},
  pages={020302},
  year={2020},
  publisher={APS},
  url={https://doi.org/10.1103/PRXQuantum.1.020302}
}

@article{pSFF,
  title={Probing many-body quantum chaos with quantum simulators},
  author={Joshi, Lata Kh and Elben, Andreas and Vikram, Amit and Vermersch, Beno{\^i}t and Galitski, Victor and Zoller, Peter},
  journal={Phys. Rev. X},
  volume={12},
  number={1},
  pages={011018},
  year={2022},
  publisher={APS},
  url={https://doi.org/10.1103/PhysRevX.12.011018}
}

@article{dynamicalqergodicity,
  title={Dynamical quantum ergodicity from energy level statistics},
  author={Vikram, Amit and Galitski, Victor},
  journal={Phys. Rev. Res.},
  volume={5},
  number={3},
  pages={033126},
  year={2023},
  publisher={APS},
  url = {https://doi.org/10.1103/PhysRevResearch.5.033126},
archivePrefix = {arXiv},
arxivId = {2205.05704}
}

@article{dynamicalqspeedlimit,
  title={Exact universal bounds on quantum dynamics and fast scrambling},
  author={Vikram, Amit and Galitski, Victor},
  journal={Phys. Rev. Lett.},
  volume={132},
  number={4},
  pages={040402},
  year={2024},
  publisher={APS},
  url={https://doi.org/10.1103/PhysRevLett.132.040402},
archivePrefix = {arXiv},
arxivId = {2212.14021}
}

@article{dynamicalqfastscrambling,
  title={Proof of a Universal Speed Limit on Fast Scrambling in Quantum Systems},
  author={Vikram, Amit and Shou, Laura and Galitski, Victor},
  journal={Phys. Rev. Lett.},
  volume={136},
  number={15},
  pages={150401},
  year={2026},
  publisher={APS},
  archivePrefix = {arXiv},
  arxivId = {2404.15403},
  url={https://doi.org/10.1103/y9z4-v641}
}

@article{QuantumRecurrences,
  title={Quantum recurrence theorem},
  author={Bocchieri, P and Loinger, A},
  journal={Phys. Rev.},
  volume={107},
  number={2},
  pages={337},
  year={1957},
  publisher={APS},
  url={https://doi.org/10.1103/PhysRev.107.337}
}

@article{BrownSusskind2,
  title={Second law of quantum complexity},
  author={Brown, Adam R and Susskind, Leonard},
  journal={Phys. Rev. D},
  volume={97},
  number={8},
  pages={086015},
  year={2018},
  publisher={APS},
  url={https://doi.org/10.1103/PhysRevD.97.086015}
}

@article{ShenkerThouless,
  title={Onset of random matrix behavior in scrambling systems},
  author={Gharibyan, Hrant and Hanada, Masanori and Shenker, Stephen H and Tezuka, Masaki},
  journal={J. High Energy Phys.},
  volume={2018},
  number={7},
  pages={1--62},
  year={2018},
  publisher={Springer},
  url={https://doi.org/10.1007/JHEP07(2018)124}
}

@article{ChanScrambling,
  title={Solution of a minimal model for many-body quantum chaos},
  author={Chan, Amos and De Luca, Andrea and Chalker, JT},
  journal={Phys. Rev. X},
  volume={8},
  number={4},
  pages={041019},
  year={2018},
  publisher={APS},
  url={https://doi.org/10.1103/PhysRevX.8.041019}
}

@article{HaydenPreskill,
  title={Black holes as mirrors: quantum information in random subsystems},
  author={Hayden, Patrick and Preskill, John},
  journal={J. High Energy Phys.},
  volume={2007},
  number={09},
  pages={120},
  year={2007},
 publisher={IOP Publishing},
  url = {https://doi.org/10.1088/1126-6708/2007/09/120}
}

@article{SekinoSusskind,
  title={Fast scramblers},
  author={Sekino, Yasuhiro and Susskind, Leonard},
  journal={J. High Energy Phys.},
  volume={2008},
  number={10},
  pages={065},
  year={2008},
  publisher={IOP Publishing},
  url={https://doi.org/10.1088/1126-6708/2008/10/065}
}

@article{LashkariFastScrambling,
  title={Towards the fast scrambling conjecture},
  author={Lashkari, Nima and Stanford, Douglas and Hastings, Matthew and Osborne, Tobias and Hayden, Patrick},
  journal={J. High Energy Phys.},
  volume={2013},
  number={4},
  pages={1--33},
  year={2013},
  publisher={Springer},
  url={https://doi.org/10.1007/JHEP04(2013)022}
}

@article{ChaosComplexityRMT,
  title={Chaos, complexity, and random matrices},
  author={Cotler, Jordan and Hunter-Jones, Nicholas and Liu, Junyu and Yoshida, Beni},
  journal={J. High Energy Phys.},
  volume={2017},
  number={11},
  pages={1--60},
  year={2017},
  publisher={Springer},
  url={https://doi.org/10.1007/JHEP11(2017)048}
}

@article{CotlerHunterJones2,
  title={Spectral decoupling in many-body quantum chaos},
  author={Cotler, Jordan and Hunter-Jones, Nicholas},
  journal={J. High Energy Phys.},
  volume={2020},
  number={12},
  pages={1--62},
  year={2020},
  publisher={Springer},
  url={https://doi.org/10.1007/JHEP12(2020)205}
}

@article{SSS,
  title={A semiclassical ramp in {SYK} and in gravity},
  author={Saad, Phil and Shenker, Stephen H and Stanford, Douglas},
  journal={arXiv preprint arXiv:1806.06840},
  year={2018},
  url={https://doi.org/10.48550/arXiv.1806.06840}
}

@article{Zelditch,
  title={Quantum ergodicity and mixing},
  author={Zelditch, Steve},
  journal={arXiv preprint math-ph/0503026},
  year={2005},
  url={https://doi.org/10.48550/arXiv.math-ph/0503026}
}

@article{BentsenGuLucasScrambling,
  title={Fast scrambling on sparse graphs},
  author={Bentsen, Gregory and Gu, Yingfei and Lucas, Andrew},
  journal={Proc. Natl. Acad. Sci. U.S.A.},
  volume={116},
  number={14},
  pages={6689--6694},
  year={2019},
  publisher={National Acad Sciences},
  url={https://doi.org/10.1073/pnas.1811033116}
}

@article{LucasEntanglementVsOTOC,
  title={Quantum many-body dynamics on the star graph},
  author={Lucas, Andrew},
  journal={arXiv preprint arXiv:1903.01468},
  year={2019},
  url={https://doi.org/10.48550/arXiv.1903.01468}
}

@article{KosProsen2018,
  title={Many-body quantum chaos: Analytic connection to random matrix theory},
  author={Kos, Pavel and Ljubotina, Marko and Prosen, Toma{\v{z}}},
  journal={Phys. Rev. X},
  volume={8},
  number={2},
  pages={021062},
  year={2018},
  publisher={APS},
  url = {https://doi.org/10.1103/PhysRevX.8.021062}
}

@article{BertiniProsen,
  title={Exact spectral form factor in a minimal model of many-body quantum chaos},
  author={Bertini, Bruno and Kos, Pavel and Prosen, Toma{\v{z}}},
  journal={Phys. Rev. Lett.},
  volume={121},
  number={26},
  pages={264101},
  year={2018},
  publisher={APS},
  url={https://doi.org/10.1103/PhysRevLett.121.264101}
}

@article{Lozej2021,
  title={Quantum chaos in triangular billiards},
  author={Lozej, {\v{C}}rt and Casati, Giulio and Prosen, Toma{\v{z}}},
  journal={Phys. Rev. Res.},
  volume={4},
  number={1},
  pages={013138},
  year={2022},
  publisher={APS},
  url={https://doi.org/10.1103/PhysRevResearch.4.013138}
}

@book{NielsenChuang,
title={Quantum Computation and Quantum Information},
publisher={Cambridge University Press},
author={Nielsen, Michael A. and Chuang, Isaac L.},
year={2010},
doi={10.1017/CBO9780511976667},
url={https://doi.org/10.1017/CBO9780511976667}
}

@article{HorodeckiEntanglementReview,
  title={Quantum entanglement},
  author={Horodecki, Ryszard and Horodecki, Pawe{\l} and Horodecki, Micha{\l} and Horodecki, Karol},
  journal={Rev. Mod. Phys.},
  volume={81},
  number={2},
  pages={865},
  year={2009},
  publisher={APS},
url={https://doi.org/10.1103/RevModPhys.81.865}
}

@article{AndreasPurityRM,
  title={Probing {R{\'e}nyi} entanglement entropy via randomized measurements},
  author={Brydges, Tiff and Elben, Andreas and Jurcevic, Petar and Vermersch, Beno{\^\i}t and Maier, Christine and Lanyon, Ben P and Zoller, Peter and Blatt, Rainer and Roos, Christian F},
  journal={Science},
  volume={364},
  number={6437},
  pages={260--263},
  year={2019},
  publisher={American Association for the Advancement of Science},
url={https://doi.org/10.1126/science.aau4963}
}

@article{pSFFexpt1,
  title={Measuring the Spectral Form Factor in Many-Body Chaotic and Localized Phases of Quantum Processors},
  author={Dong, Hang and Zhang, Pengfei and Da{\u{g}}, Ceren B and Gao, Yu and Wang, Ning and Deng, Jinfeng and Zhang, Xu and Chen, Jiachen and Xu, Shibo and Wang, Ke and others},
  journal={Phys. Rev. Lett.},
  volume={134},
  number={1},
  pages={010402},
  year={2025},
  publisher={APS},
  url={https://doi.org/10.1103/PhysRevLett.134.010402}
}

@article{bakeranomalieserg,
  title={Spectral anomalies and broken symmetries in maximally chaotic quantum maps},
  author={Shou, Laura and Vikram, Amit and Galitski, Victor},
  journal={SciPost Phys.},
  volume={18},
  pages={151},
  year={2025},
  publisher={SciPost},
  doi={10.21468/SciPostPhys.18.5.151},
  url={https://doi.org/10.21468/SciPostPhys.18.5.151}
}

@Book{Sinai1976,
  author    = {Sinai, Yakov G.},
  year      = {1977},
  title     = {Introduction to ergodic theory},
  isbn      = {978-0691081823},
  publisher = {Princeton University Press},
  volume    = {18},
}

@Book{HalmosErgodic,
  author    = {Halmos, Paul R},
  year      = {2017},
  title     = {Lectures on ergodic theory},
  isbn      = {9780486814896},
  publisher = {Dover Publications},
}

@Book{SinaiCornfeld,
  author    = {Cornfeld, I. P. and Fomin, S. V. and Sinai, Yakov G.},
  year      = {1982},
  title     = {Ergodic theory},
  isbn      = {978-1-4615-6927-5},
  publisher = {Springer-Verlag New York},
  url       = {https://doi.org/10.1007/978-1-4615-6927-5},
}

@Article{PlatoErgodic,
  author       = {Frigg, Roman and Berkovitz, Joseph and Kronz, Fred},
  journal = {The Stanford Encyclopedia of Philosophy, Fall 2020 Edition},
  title        = {The Ergodic Hierarchy},
  editor       = {Edward N. Zalta},
  url          = {https://plato.stanford.edu/archives/fall2020/entries/ergodic-hierarchy/},
year={2020}
}

@Article{BerryTabor,
  author       = {Berry, M. V. and Tabor, M.},
  year         = {1977},
  journal = {Proc. Roy. Soc. Lond. A},
  title        = {{Level clustering in the regular spectrum}},
  pages        = {375--394},
  url          = {https://royalsocietypublishing.org/doi/10.1098/rspa.1977.0140},
  volume       = {356},
}

@Article{BerrySpectralRigidity,
  author       = {Berry, Michael Victor},
  year         = {1985},
  journal = {Proc. Roy. Soc. Lond. A},
  title        = {Semiclassical theory of spectral rigidity},
  number       = {1819},
  pages        = {229--251},
  url          = {https://doi.org/10.1098/rspa.1985.0078},
  volume       = {400},
  publisher    = {The Royal Society London},
}

@InCollection{Berry227,
  author    = {Berry, Michael V},
  booktitle = {Les Houches Lecture series {LII} (1989)},
  year      = {1991},
  title     = {Some quantum-to-classical asymptotics},
  editor    = {Giannoni, M- J and Voros, A and Zinn-Justin, J},
  isbn      = {978-0444892775},
  pages     = {251--304},
  publisher = {Elsevier Science Publishers B.V.},
  url       = {https://michaelberryphysics.files.wordpress.com/2013/07/berry227.pdf},
}

@Article{ThoulessRelaxation,
  author       = {Schiulaz, Mauro and Torres-Herrera, E Jonathan and Santos, Lea F},
  year         = {2019},
  journal = {Phys. Rev. B},
  title        = {Thouless and relaxation time scales in many-body quantum systems},
  number       = {17},
  pages        = {174313},
  url          = {https://doi.org/10.1103/PhysRevB.99.174313},
  volume       = {99},
  publisher    = {APS},
}

@Article{DKchaos,
  author       = {McDonald, Steven W and Kaufman, Allan N},
  year         = {1979},
  journal = {Phys. Rev. Lett.},
  title        = {Spectrum and eigenfunctions for a {Hamiltonian} with stochastic trajectories},
  number       = {18},
  pages        = {1189},
  url          = {https://doi.org/10.1103/PhysRevLett.42.1189},
  volume       = {42},
  publisher    = {APS}
}

@Article{CGV,
  author       = {Casati, G and Valz-Gris, F and Guarnieri, I},
  year         = {1980},
  journal = {Lett. Nuovo Cimento},
  title        = {On the connection between quantization of nonintegrable systems and statistical theory of spectra},
  number       = {8},
  pages        = {279--282},
  volume       = {28}
}

@Article{BerryStadium,
  author       = {Berry, Michael V},
  year         = {1981},
  journal = {Ann. Phys.},
  title        = {Quantizing a classically ergodic system: {Sinai}'s billiard and the {KKR} method},
  number       = {1},
  pages        = {163--216},
  url          = {https://doi.org/10.1016/0003-4916(81)90189-5},
  volume       = {131},
  publisher    = {Elsevier}
}

@Article{BGS,
  author       = {Bohigas, Oriol and Giannoni, Marie-Joya and Schmit, Charles},
  year         = {1984},
  journal = {Phys. Rev. Lett.},
  title        = {Characterization of chaotic quantum spectra and universality of level fluctuation laws},
  number       = {1},
  pages        = {1},
  url          = {https://doi.org/10.1103/PhysRevLett.52.1},
  volume       = {52},
  publisher    = {APS}
}

@Article{WuErgMix,
  author       = {Zhang, Dongliang and Quan, H T and Wu, Biao},
  year         = {2016},
  journal = {Phys. Rev. E},
  title        = {Ergodicity and mixing in quantum dynamics},
  number       = {2},
  pages        = {022150},
  url          = {https://doi.org/10.1103/PhysRevE.94.022150},
  volume       = {94},
  publisher    = {APS},
}

@Article{CasatiWang,
  author       = {Wang, Jiaozi and Benenti, Giuliano and Casati, Giulio and Wang, Wen-ge},
  year         = {2022},
  journal = {J. Phys. A: Math. Theor},
  title        = {Statistical and dynamical properties of the quantum triangle map},
  pages        = {234002},
  url          = {https://doi.org/10.1088/1751-8121/ac6a93},
  volume       = {55},
}

@InProceedings{Anantharaman,
  author       = {Anantharaman, Nalini},
  booktitle    = {{Proceedings of the International Congress of Mathematicians: Rio de Janeiro 2018}},
  year         = {2018},
  title        = {Delocalization of {Schr{\"o}dinger} eigenfunctions},
  organization = {World Scientific},
  pages        = {341--375},
  url          = {https://doi.org/10.1142/9789813272880_0016},
}

@Article{GiraudWDwithoutPO,
  author       = {Giraud, Olivier and Marklof, Jens and O'Keefe, Stephen},
  year         = {2004},
  journal = {J. Phys. A: Math. Gen.},
  title        = {Intermediate statistics in quantum maps},
  number       = {28},
  pages        = {L303},
  url          = {https://doi.org/10.1088/0305-4470/37/28/L01},
  volume       = {37},
  publisher    = {IOP Publishing},
}

@article{refRampPlateau2,
  title={Spectral form factors and late time quantum chaos},
  author={Liu, Junyu},
  journal={Phys. Rev. D.},
  volume={98},
  number={8},
  pages={086026},
  year={2018},
  publisher={APS},
  url={https://doi.org/10.1103/PhysRevD.98.086026}
}

@article{keating1991cat,
  title={The cat maps: quantum mechanics and classical motion},
  author={Keating, Jonathan P},
  journal={Nonlinearity},
  volume={4},
  number={2},
  pages={309},
  year={1991},
  publisher={IOP Publishing},
  url={https://doi.org/10.1088/0951-7715/4/2/006}
}

@article{BraunHaakearithmetic,
  title={Level statistics in arithmetical and pseudo-arithmetical chaos},
  author={Braun, Petr and Haake, Fritz},
  journal={J. Phys. A: Math. Theor.},
  volume={43},
  number={26},
  pages={262001},
  year={2010},
  publisher={IOP Publishing},
  url={https://doi.org/10.1088/1751-8113/43/26/262001}
}

@article{BGGSarithmetic,
  author = {Bogomolny, E.B. and Georgeot, B. and Giannoni, M.-J. and Schmit, C.},
  title = {Arithmetical chaos},
  journal = {Phys. Rep.},
  fjournal = {Physics Reports},
  volume = {291},
  number = {5--6},
  pages = {219--324},
  month = {December},
  year = {1997},
  publisher = {Elsevier},
  address = {Amsterdam, Netherlands},
  doi = {10.1016/S0370-1573(97)00016-1},
}

@article{BV,
    AUTHOR = {Balazs, N. L. and Voros, A.},
     TITLE = {The quantized baker's transformation},
   JOURNAL = {Ann. Physics},
  FJOURNAL = {Annals of Physics},
    VOLUME = {190},
      YEAR = {1989},
    NUMBER = {1},
     PAGES = {1--31},
doi={10.1209/0295-5075/4/10/001},
}

@article{Saraceno,
    AUTHOR = {Saraceno, M.},
     TITLE = {Classical structures in the quantized baker transformation},
   JOURNAL = {Ann. Physics},
  FJOURNAL = {Annals of Physics},
    VOLUME = {199},
      YEAR = {1990},
    NUMBER = {1},
     PAGES = {37--60},
doi={10.1016/0003-4916(90)90367-W},
}

@article{ThermalNEEs,
  title = {Quantum Ergodicity in the Many-Body Localization Problem},
  author = {Monteiro, Felipe and Tezuka, Masaki and Altland, Alexander and Huse, David A. and Micklitz, Tobias},
  journal = {Phys. Rev. Lett.},
  volume = {127},
  issue = {3},
  pages = {030601},
  numpages = {6},
  year = {2021},
  month = {Jul},
  publisher = {American Physical Society},
  doi = {10.1103/PhysRevLett.127.030601},
  url = {https://link.aps.org/doi/10.1103/PhysRevLett.127.030601}
}

@article{Suntajs2019,
  title = {Quantum chaos challenges many-body localization},
  author = {\ifmmode \check{S}\else \v{S}\fi{}untajs, Jan and Bon\ifmmode \check{c}\else \v{c}\fi{}a, Janez and Prosen, Toma\v{z} and Vidmar, Lev},
  journal = {Phys. Rev. E},
  volume = {102},
  issue = {6},
  pages = {062144},
  numpages = {12},
  year = {2020},
  month = {Dec},
  publisher = {American Physical Society},
  doi = {10.1103/PhysRevE.102.062144},
  url = {https://link.aps.org/doi/10.1103/PhysRevE.102.062144}
}

@article{lu2019renyi,
  title = {Renyi entropy of chaotic eigenstates},
  author = {Lu, Tsung-Cheng and Grover, Tarun},
  journal = {Phys. Rev. E},
  volume = {99},
  issue = {3},
  pages = {032111},
  numpages = {20},
  year = {2019},
  month = {Mar},
  publisher = {American Physical Society},
  doi = {10.1103/PhysRevE.99.032111},
  url = {https://link.aps.org/doi/10.1103/PhysRevE.99.032111}
}

@article{RudnickWhatIsQChaos,
  title={What is quantum chaos},
  author={Rudnick, Ze’ev},
  journal={Notices of the AMS},
  volume={55},
  number={1},
  pages={32--34},
  year={2008},
  url={https://www.ams.org/notices/200801/tx080100032p.pdf}
}

@article{JTreconstruction,
author = {Akers, Chris and Lucas, Andrew and Vikram, Amit},
title={On the reconstruction map in {JT} gravity},
journal={J. High Energy Phys.},
  volume={2025},
  number={12},
  pages={045},
  year={2025},
  publisher={Springer},
  url={https://doi.org/10.1007/JHEP12(2025)045}
}

@book{KhinchinStatMech,
  title={Mathematical foundations of statistical mechanics},
  author={Khinchin, Aleksandr Yakovlevich},
  year={1949},
  publisher={Courier Corporation},
  isbn={9780486601472}
}

@article{JensenShankarETH,
  title={Statistical behavior in deterministic quantum systems with few degrees of freedom},
  author={Jensen, R V and Shankar, R},
  journal={Phys. Rev. Lett.},
  volume={54},
  number={17},
  pages={1879},
  year={1985},
  publisher={APS},
  url={https://doi.org/10.1103/PhysRevLett.54.1879}
}

@article{AdwaySantos,
  title={Proposal for many-body quantum chaos detection},
  author={Das, Adway Kumar and Cianci, Cameron and Cabral, Delmar GA and Zarate-Herrada, David A and Pinney, Patrick and Pilatowsky-Cameo, Sa{\'u}l and Matsoukas-Roubeas, Apollonas S and Batista, Victor S and del Campo, Adolfo and Torres-Herrera, E Jonathan and others},
  journal={Phys. Rev. Res.},
  volume={7},
  number={1},
  pages={013181},
  year={2025},
  publisher={APS},
  url={https://doi.org/10.1103/PhysRevResearch.7.013181}
}

@article{MaganWu,
  title={Two types of quantum chaos: testing the limits of the {Bohigas-Giannoni-Schmit} conjecture},
  author={Magan, Javier M and Wu, Qingyue},
  journal={arXiv preprint arXiv:2411.08186},
  year={2024},
  url={https://doi.org/10.48550/arXiv.2411.08186}
}

@article{SusskindCV,
  title={Addendum to Computational Complexity and Black Hole Horizons},
  author={Susskind, Leonard},
  journal={arXiv preprint arXiv:1403.5695},
  year={2014},
  url={https://doi.org/10.48550/arXiv.1403.5695}
}

@article{Shnirelman,
  title={Ergodic properties of eigenfunctions},
  author={Shnirelman, Alexander I},
  journal={Uspekhi Mat. Nauk},
  volume={29},
  number={6},
  pages={181--182},
  year={1974}
}

@article{CdV,
  title={Ergodicit{\'e} et fonctions propres du laplacien},
  author={Colin de Verdi{\`e}re, Y},
  journal={Commun. Math. Phys.},
  volume={102},
  pages={497--502},
  year={1985},
  publisher={Springer},
  url={https://doi.org/10.1007/BF01209296}
}

@article{ZelditchOG,
  title={Uniform distribution of eigenfunctions on compact hyperbolic surfaces},
  author={Zelditch, Steven},
  year={1987},
  journal = {Duke Mathematical Journal},
  volume={55},
  pages={919-941},
  url={https://doi.org/10.1215/S0012-7094-87-05546-3}
}

@incollection{Sunada,
  title={Quantum ergodicity},
  author={Sunada, Toshikazu},
  booktitle={Progress in inverse spectral geometry},
  pages={175--196},
  year={1997},
  publisher={Springer},
  url={https://doi.org/10.1007/978-3-0348-8938-4_10}
}

@article{ZelditchTransition,
  title={Quantum transition amplitudes for ergodic and for completely integrable systems},
  author={Zelditch, Steven},
  journal={J. Funct. Anal.},
  volume={94},
  number={2},
  pages={415--436},
  year={1990},
  publisher={Academic Press},
  url={https://doi.org/10.1016/0022-1236(90)90021-C}
}

@article{ZelditchMixing,
  title={Quantum mixing},
  author={Zelditch, Steven},
  journal={J. Funct. Anal.},
  volume={140},
  number={1},
  pages={68--86},
  year={1996},
  publisher={Elsevier},
  url={https://doi.org/10.1006/jfan.1996.0098}
}

@book{ByronFuller,
  title={Mathematics of classical and quantum physics},
  author={Byron, Frederick W and Fuller, Robert W},
  year={2012},
  publisher={Dover Publications},
  isbn={9780486135069}
}

@article{spreadcomplexity,
  title={Quantum chaos and the complexity of spread of states},
  author={Balasubramanian, Vijay and Caputa, Pawel and Magan, Javier M and Wu, Qingyue},
  journal={Phys. Rev. D},
  volume={106},
  number={4},
  pages={046007},
  year={2022},
  publisher={APS},
  url={https://doi.org/10.1103/PhysRevD.106.046007},
arxivPrefix={arXiv},
eprint={2202.06957}
}

@article{WilkieBrumerReturnProbabilities,
  title={Time-dependent manifestations of quantum chaos},
  author={Wilkie, Joshua and Brumer, Paul},
  journal={Phys. Rev. Lett.},
  volume={67},
  number={10},
  pages={1185},
  year={1991},
  publisher={APS},
url={https://doi.org/10.1103/PhysRevLett.67.1185}
}

@article{KrylovReview,
  title={Quantum dynamics in {Krylov} space: Methods and applications},
  author={Nandy, Pratik and Matsoukas-Roubeas, Apollonas S and Mart{\'\i}nez-Azcona, Pablo and Dymarsky, Anatoly and del Campo, Adolfo},
  journal={Physics Reports},
  volume={1125},
  pages={1--82},
  year={2025},
  publisher={Elsevier},
url={https://doi.org/10.1016/j.physrep.2025.05.001},
archivePrefix={arXiv},
eprint={2405.09628}
}

@article{ComplexityAnything,
  title={Does complexity equal anything?},
  author={Belin, Alexandre and Myers, Robert C and Ruan, Shan-Ming and S{\'a}rosi, G{\'a}bor and Speranza, Antony J},
  journal={Phys. Rev. Lett.},
  volume={128},
  number={8},
  pages={081602},
  year={2022},
  publisher={APS},
url={https://doi.org/10.1103/PhysRevLett.128.081602},
archivePrefix={arXiv},
eprint={2111.02429}
}

@article{dynamicalqthermalization,
  title={Bypassing eigenstate thermalization with experimentally accessible quantum dynamics},
  author={Vikram, Amit},
  journal={SciPost Phys.},
  volume={21},
  pages={029},
  year={2026},
  publisher={SciPost},
  archivePrefix = {arXiv},
  arxivId = {2503.07729},
url={https://doi.org/10.21468/SciPostPhys.21.1.029}
}

@article{dynamicalpurestatethermalization,
  title={Provable quantum thermalization without statistical averages},
  author={Vikram, Amit},
  journal={arXiv preprint arXiv:2604.02417},
  year={2026},
  url={https://doi.org/10.48550/arXiv.2604.02417}
}

@article{dynamicalqaperiodicity,
  title={Aperiodicity is sufficient for macroscopic thermalization},
  author={Vikram, Amit},
  journal={arXiv preprint arXiv:2608.13462},
  year={2026},
  url={https://doi.org/10.48550/arXiv.2608.13462}
}

@article{QuantumOperationsReview,
  title={Higher-Order Quantum Operations},
  author={Taranto, Philip and Milz, Simon and Murao, Mio and Quintino, Marco T{\'u}lio and Modi, Kavan},
  journal={arXiv preprint arXiv:2503.09693},
  year={2025},
url={https://doi.org/10.48550/arXiv.2503.09693}
}

@article{TorresHerreraSantos,
  title={Dynamical manifestations of quantum chaos: correlation hole and bulge},
  author={Torres-Herrera, EJ and Santos, Lea F},
  journal={Philos. Trans. R. Soc. A},
  volume={375},
  number={2108},
  pages={20160434},
  year={2017},
  publisher={The Royal Society Publishing},
url={https://doi.org/10.1098/rsta.2016.0434}
}

@book{GaussianMaxdist1,
  title={Concentration inequalities and model selection: Ecole d'Et{\'e} de Probabilit{\'e}s de Saint-Flour XXXIII-2003},
  author={Massart, Pascal},
  year={2007},
  publisher={Springer},
pages={17, Lemma 2.3},
url={https://doi.org/10.1007/978-3-540-48503-2},
doi={10.1007/978-3-540-48503-2}
}

@Inbook{GaussianMaxdist2,
title="Gaussian Inequalities",
bookTitle="Random Fields and Geometry",
year="2007",
publisher="Springer New York",
address="New York, NY",
pages="49--64",
author = {Adler, Robert J and Taylor, Jonathan E},
isbn="978-0-387-48116-6",
doi="10.1007/978-0-387-48116-6_2",
url="https://doi.org/10.1007/978-0-387-48116-6_2"
}

@article{GaussianMaxdist3,
  title={Optimal non-asymptotic lower bound on the minimax regret of learning with expert advice},
  author={Orabona, Francesco and Pal, David},
  journal={arXiv preprint arXiv:1511.02176},
  year={2015},
url={https://doi.org/10.48550/arXiv.1511.02176}
}

@article{EPR,
  title={Can quantum-mechanical description of physical reality be considered complete?},
  author={Einstein, Albert and Podolsky, Boris and Rosen, Nathan},
  journal={Phys. Rev.},
  volume={47},
  number={10},
  pages={777},
  year={1935},
  publisher={APS},
url={https://doi.org/10.1103/PhysRev.47.777}
}

@article{BennettTeleportation,
  title={Teleporting an unknown quantum state via dual classical and {Einstein-Podolsky-Rosen} channels},
  author={Bennett, Charles H and Brassard, Gilles and Cr{\'e}peau, Claude and Jozsa, Richard and Peres, Asher and Wootters, William K},
  journal={Phys. Rev. Lett.},
  volume={70},
  number={13},
  pages={1895},
  year={1993},
  publisher={APS},
url={https://doi.org/10.1103/PhysRevLett.70.1895}
}

@article{TeleportationReview1,
  title={Advances in quantum teleportation},
  author={Pirandola, Stefano and Eisert, Jens and Weedbrook, Christian and Furusawa, Akira and Braunstein, Samuel L},
  journal={Nature Photon.},
  volume={9},
  number={10},
  pages={641--652},
  year={2015},
  publisher={Nature Publishing Group UK London},
url={https://doi.org/10.1038/nphoton.2015.154}
}

@article{TeleportationReview2,
  title={Progress in quantum teleportation},
  author={Hu, Xiao-Min and Guo, Yu and Liu, Bi-Heng and Li, Chuan-Feng and Guo, Guang-Can},
  journal={Nat. Rev. Phys.},
  volume={5},
  number={6},
  pages={339--353},
  year={2023},
  publisher={Nature Publishing Group UK London},
url={https://doi.org/10.1038/s42254-023-00588-x}
}

@article{KCRTeleportation,
  title={Generating {Einstein-Podolsky-Rosen} correlations for teleporting collective spin states in a two-dimensional trapped ion crystal},
  author={Khan, Muhammad Miskeen and Chaparro, Edwin and Sundar, Bhuvanesh and Carter, Allison and Bollinger, John and Molmer, Klaus and Rey, Ana Maria},
  journal={Phys. Rev. Res.},
  volume={7},
  number={2},
  pages={L022019},
  year={2025},
  publisher={APS},
url={https://doi.org/10.1103/PhysRevResearch.7.L022019}
}

@article{QNDspin,
  title={Quantum measurement of a mesoscopic spin ensemble},
  author={Giedke, Geza and Taylor, Jacob M and D’Alessandro, Domenico and Lukin, Mikhail D and Imamo{\u{g}}lu, A},
  journal={Phys. Rev. A},
  volume={74},
  number={3},
  pages={032316},
  year={2006},
  publisher={APS},
url={https://doi.org/10.1103/PhysRevA.74.032316}
}

@article{QNDspinQDOT,
  title={Controlling a mesoscopic spin environment by quantum bit manipulation},
  author={Taylor, JM and Imamoglu, Atac and Lukin, MD},
  journal={Phys. Rev. Lett.},
  volume={91},
  number={24},
  pages={246802},
  year={2003},
  publisher={APS},
url={https://doi.org/10.1103/PhysRevLett.91.246802}
}

@article{KrylovOG,
  title={De la r\'{e}solution num\'{e}rique de l'\'{e}quation servant \`{a} d\'{e}terminer dans des questions de m\'{e}canique appliqu\'{e}e les fr\'{e}quences de petites oscillations des syst\`{e}mes mat\'{e}riels},
  author={Krylov, Aleksei Nikolaevich},
  journal={Izv. Math.},
  number={4},
  pages={491--539},
  year={1931},
  publisher={Russian Academy of Sciences, Steklov Mathematical Institute},
url={http://mi.mathnet.ru/eng/im5215}
}

@book{KrylovBook,
  title={Krylov subspace methods: principles and analysis},
  author={Liesen, J{\"o}rg and Strakos, Zdenek},
  year={2013},
  publisher={Oxford University Press},
isbn={9780199655410},
url={https://doi.org/10.1093/acprof:oso/9780199655410.001.0001}
}

@inproceedings{SchrodingerHJSW,
  title={Discussion of probability relations between separated systems},
  author={Schr{\"o}dinger, Erwin},
  booktitle={Math. Proc. Camb. Philos. Soc.},
  volume={31},
  number={4},
  pages={555--563},
  year={1935},
  organization={Cambridge University Press},
url={https://doi.org/10.1017/S0305004100013554}
}

@article{HJSW2,
  title={A complete classification of quantum ensembles having a given density matrix},
  author={Hughston, Lane P and Jozsa, Richard and Wootters, William K},
  journal={Phys. Lett. A},
  volume={183},
  number={1},
  pages={14--18},
  year={1993},
  publisher={Elsevier},
url={https://doi.org/10.1016/0375-9601(93)90880-9}
}

@article{coherentGibbs,
  title={Quantum speed limits on operator flows and correlation functions},
  author={Carabba, Nicoletta and H{\"o}rnedal, Niklas and del Campo, Adolfo},
  journal={Quantum},
  volume={6},
  pages={884},
  year={2022},
  publisher={Verein zur F{\"o}rderung des Open Access Publizierens in den Quantenwissenschaften},
url={https://doi.org/10.22331/q-2022-12-22-884}
}

@article{GorinSeligman2001,
  title={A random matrix approach to decoherence},
  author={Gorin, Thomas and Seligman, Thomas H},
  journal={J. Opt. B: Quantum Semiclass. Opt.},
  volume={4},
  number={4},
  pages={S386},
  year={2002},
  publisher={IOP Publishing},
url={https://doi.org/10.1088/1464-4266/4/4/325}
}

@article{VinayakZnidaric,
  title={Subsystem dynamics under random {Hamiltonian} evolution},
  author={Vinayak and {\v{Z}}nidari{\v{c}}, Marko},
  journal={J. Phys. A.: Math. Theor.},
  volume={45},
  number={12},
  pages={125204},
  year={2012},
  publisher={IOP Publishing},
url={https://doi.org/10.1088/1751-8113/45/12/125204}
}

@article{BhatiaDavis,
  title={A better bound on the variance},
  author={Bhatia, Rajendra and Davis, Chandler},
  journal={The American Mathematical Monthly},
  volume={107},
  number={4},
  pages={353--357},
  year={2000},
  publisher={Taylor \& Francis},
url={https://doi.org/10.1080/00029890.2000.12005203}
}

@article{LiuXiaoBalents_SYKEntanglement,
  title={Quantum entanglement of the {Sachdev-Ye-Kitaev} models},
  author={Liu, Chunxiao and Chen, Xiao and Balents, Leon},
  journal={Phys. Rev. B},
  volume={97},
  number={24},
  pages={245126},
  year={2018},
  publisher={APS},
url={https://doi.org/10.1103/PhysRevB.97.245126}
}

@article{RandomCircuitEntanglement,
  title={Quantum entanglement in random physical states},
  author={Hamma, Alioscia and Santra, Siddhartha and Zanardi, Paolo},
  journal={Phys. Rev. Lett.},
  volume={109},
  number={4},
  pages={040502},
  year={2012},
  publisher={APS},
url={https://doi.org/10.1103/PhysRevLett.109.040502}
}

@article{PageSubsystem,
  title={Average entropy of a subsystem},
  author={Page, Don N},
  journal={Phys. Rev. Lett.},
  volume={71},
  number={9},
  pages={1291},
  year={1993},
  publisher={APS},
url={https://doi.org/10.1103/PhysRevLett.71.1291}
}

@article{HighDimEntanglementReview,
  title={Advances in high-dimensional quantum entanglement},
  author={Erhard, Manuel and Krenn, Mario and Zeilinger, Anton},
  journal={Nat. Rev. Phys.},
  volume={2},
  number={7},
  pages={365--381},
  year={2020},
  publisher={Nature Publishing Group UK London},
url={https://doi.org/10.1038/s42254-020-0193-5}
}

@article{HighDimEntanglement1,
  title={Experimental high-dimensional quantum teleportation},
  author={Hu, Xiao-Min and Zhang, Chao and Liu, Bi-Heng and Cai, Yu and Ye, Xiang-Jun and Guo, Yu and Xing, Wen-Bo and Huang, Cen-Xiao and Huang, Yun-Feng and Li, Chuan-Feng and others},
  journal={Phys. Rev. Lett.},
  volume={125},
  number={23},
  pages={230501},
  year={2020},
  publisher={APS},
url={https://doi.org/10.1103/PhysRevLett.125.230501}
}

@article{HighDimEntanglement2,
  title={Quantum teleportation in high dimensions},
  author={Luo, Yi-Han and Zhong, Han-Sen and Erhard, Manuel and Wang, Xi-Lin and Peng, Li-Chao and Krenn, Mario and Jiang, Xiao and Li, Li and Liu, Nai-Le and Lu, Chao-Yang and others},
  journal={Phys. Rev. Lett.},
  volume={123},
  number={7},
  pages={070505},
  year={2019},
  publisher={APS},
url={https://doi.org/10.1103/PhysRevLett.123.070505}
}

@article{BerryChaology,
  title={{The Bakerian Lecture, 1987. Quantum chaology}},
  author={Berry, Michael Victor},
  journal={Proc. Roy. Soc. Lond. A},
  volume={413},
  number={1844},
  pages={183--198},
  year={1987},
  publisher={The Royal Society London},
url={https://doi.org/10.1098/rspa.1987.0109}
}

@article{BerryChaologyNotChaos,
  title={Quantum chaology, not quantum chaos},
  author={Berry, Michael},
  journal={Phys. Scr.},
  volume={40},
  number={3},
  pages={335},
  year={1989},
  publisher={IOP Publishing},
url={https://doi.org/10.1088/0031-8949/40/3/013}
}

@article{TimeEstimationPageWootters,
  title={Evolution without evolution: Dynamics described by stationary observables},
  author={Page, Don N and Wootters, William K},
  journal={Phys. Rev. D},
  volume={27},
  number={12},
  pages={2885},
  year={1983},
  publisher={APS},
  url={https://doi.org/10.1103/PhysRevD.27.2885}
}

@article{TimeEstimation1,
  title={Quantum time},
  author={Giovannetti, Vittorio and Lloyd, Seth and Maccone, Lorenzo},
  journal={Phys. Rev. D},
  volume={92},
  number={4},
  pages={045033},
  year={2015},
  publisher={APS},
  url={https://doi.org/10.1103/PhysRevD.92.045033}
}

@article{TimeEstimation2,
  title={System-time entanglement in a discrete-time model},
  author={Boette, A and Rossignoli, R and Gigena, Nicol{\'a}s and Cerezo, M},
  journal={Phys. Rev. A},
  volume={93},
  number={6},
  pages={062127},
  year={2016},
  publisher={APS},
  url={https://doi.org/10.1103/PhysRevA.93.062127}
}

@article{PechukasWhatIsQuantumChaos,
  title={Kolmogorov entropy and ``quantum chaos''},
  author={Pechukas, Philip},
  journal={J. Phys. Chem.},
  volume={86},
  number={12},
  pages={2239--2243},
  year={1982},
  publisher={ACS Publications},
  url={https://doi.org/10.1021/j100209a019}
}

@article{scramblingSWAP,
  title={Bidirectional teleportation using scrambling dynamics: a practical protocol},
  author={Vikram, Amit and Chaparro, Edwin and Khan, Muhammad Miskeen and Lucas, Andrew and Akers, Chris and Rey, Ana Maria},
  journal={Phys. Rev. Lett.},
  year={to appear},
  archivePrefix = {arXiv},
  arxivId = {2601.15536},
  url={https://doi.org/10.1103/9msw-jw53}
}

@article{BricmontErgodicityCritique,
  title={Science of chaos or chaos in science?},
  author={Bricmont, Jean},
  journal={arXiv preprint chao-dyn/9603009},
  year={1996},
  url={https://doi.org/10.48550/arXiv.chao-dyn/9603009}
}

@article{GoldsteinErgodicityCritique,
  title={Boltzmann’s approach to statistical mechanics},
  author={Goldstein, Sheldon},
  journal={arXiv preprint cond-mat/0105242},
  year={2001},
  url={https://doi.org/10.48550/arXiv.cond-mat/0105242}
}

@article{LebowitzAPS2021,
  title={Microscopic Origins of Macroscopic Behavior},
  author={Lebowitz, Joel L},
  journal={arXiv preprint arXiv:2105.03470},
  year={2021},
  url={https://doi.org/10.48550/arXiv.2105.03470}
}

@article{YoshidaKitaev,
  title={Efficient decoding for the {Hayden-Preskill} protocol},
  author={Yoshida, Beni and Kitaev, Alexei},
  journal={arXiv preprint arXiv:1710.03363},
  year={2017},
url={https://doi.org/10.48550/arXiv.1710.03363}
}

@article{YoshidaYao,
  title={Disentangling scrambling and decoherence via quantum teleportation},
  author={Yoshida, Beni and Yao, Norman Y},
  journal={Phys. Rev. X},
  volume={9},
  number={1},
  pages={011006},
  year={2019},
  publisher={APS},
url={https://doi.org/10.1103/PhysRevX.9.011006}
}

@article{PeresErgodic,
  title={Ergodicity and mixing in quantum theory. {I}},
  author={Peres, Asher},
  journal={Phys. Rev. A},
  volume={30},
  number={1},
  pages={504},
  year={1984},
  publisher={APS},
  url={https://doi.org/10.1103/PhysRevA.30.504}
}

@article{BKPRMTSFFCMP,
  title={Random matrix spectral form factor of dual-unitary quantum circuits},
  author={Bertini, Bruno and Kos, Pavel and Prosen, Toma{\v{z}}},
  journal={Commun. in Math. Phys.},
  volume={387},
  number={1},
  pages={597--620},
  year={2021},
  publisher={Springer},
  url={https://doi.org/10.1007/s00220-021-04139-2}
}

\end{document}